\documentclass[prd,10pt,notitlepage,nofootinbib]{revtex4}

\usepackage{graphicx}
\usepackage{amsmath}
\usepackage{amssymb}
\usepackage{dcolumn}
\usepackage{bm}
\usepackage{color}

\newcommand \beq{\begin{eqnarray}}
\newcommand \eeq{\end{eqnarray}}

\newcommand \be{\begin{equation}}
\newcommand \ee{\end{equation}}

\newcommand{\rt}[1]{{}}

\begin{document}

\title{Broken phase scalar effective potential\\ and $\Phi$-derivable approximations}

\author{Urko Reinosa}
\email{reinosa@cpht.polytechnique.fr}
\affiliation{Centre de Physique Th{\'e}orique, Ecole Polytechnique, CNRS, 91128 Palaiseau Cedex, France.}

\author{Zsolt Sz{\'e}p}
\email{szepzs@achilles.elte.hu}
\affiliation{Centre de Physique Th{\'e}orique, Ecole Polytechnique, CNRS, 91128 Palaiseau Cedex, France.}

\date{\today}

\begin{abstract}
We study the effective potential of a real scalar $\varphi^4$ theory as a function of the temperature $T$ within the simplest $\Phi$-derivable approximation, namely the Hartree approximation. We apply renormalization at a ``high'' temperature $T_\star$ where the theory is required to be in its symmetric phase and study how the effective potential evolves as the temperature is lowered down to $T=0$. In particular, we prove analytically that no second order phase transition can occur in this particular approximation of the theory, in agreement with earlier studies based on the numerical evaluation or the high temperature expansion of the effective potential. This work is also an opportunity to illustrate certain issues on the renormalization of $\Phi$-derivable approximations at finite temperature and non-vanishing field expectation value, and to introduce new computational techniques which might also prove useful when dealing with higher order approximations.
\end{abstract}

\maketitle 

\section{Introduction}

The two-particle-irreducible (2PI) effective action \cite{LW,LY,Baym1,Baym2,Cornwall:vz} offers a general and systematically improvable approach for resumming infinite classes of Feynman diagrams of a given quantum field theory. One of its most compelling aspects is that it appears to be applicable to a wide variety of situations ranging from out-of-equilibrium settings to finite temperature calculations in equilibrium. In fact, the first orders of approximation of the 2PI effective action seem to capture already many interesting features of quantum field theories. In the case of scalar fields for instance, including the first non-local, field independent contribution to the 2PI effective action is enough to obtain a controlled (non-secular) time evolution which shows thermalization at late times, at least at the level of the two-point function \cite{Berges:2000ur}. In equilibrium, the same orders of approximation lead to a rather good convergence of some thermodynamic quantities, such as the pressure \cite{Berges:2004hn}, indicating that part of the infrared physics is properly taken into account. These observations are not limited to scalar theories but concern also theories coupling scalar and fermionic degrees of freedom (out-of-equilibrium) \cite{Berges:2002wr} or gauge theories (at finite temperature) \cite{Blaizot:2000fc,Blaizot:2005wr}. Some more genuine non-perturbative effects can be captured by combining the 2PI effective action with a large-N type expansion \cite{Berges:2001fi,Aarts:2002dj}. In the non-equilibrium context the 2PI 1/N expansion at next-to-leading order proved to be particularly fruitful for the study of a variety of problems, such that thermalization \cite{Berges:2001fi,Aarts:2002dj,Cooper:2002qd}, preheating \cite{Berges:2002cz,Arrizabalaga:2004iw}, transport coefficients \cite{Aarts:2003bk,Aarts:2004sd,Aarts:2005vc}, non-thermal fixed points \cite{Berges:2008wm}, decoherence \cite{Giraud:2009tn} and topological defect formation \cite{Berges:2010nk}. In equilibrium, the 2PI formalism was applied to phenomenological studies in various approximations, see \cite{Petropoulos:2004bt} and references therein as well as \cite{Andersen:2008qk,Roder:2005vt}. Of course, one can easily imagine that certain non-perturbative features are beyond reach within the 2PI framework or its present approximations. However, higher order approximations of the 2PI effective action or its generalization to 3PI, 4PI, ..., NPI effective actions \cite{DeDom}, although numerically unaffordable at present, could allow to capture an increasing number of such effects in the future \cite{Carrington:2004sn,Berges:2004pu,Carrington:2010qq}. Some investigations in these directions can be found in \cite{Aarts:2006cv,Carrington:2009zz,Carrington:2009kh}.\\

Concomitantly with the increasing number of applications of the 2PI effective action, new insight has been gained on technical aspects regarding its renormalization. The systematic renormalization of approximations based on the 2PI effective action was first understood in the case of scalar theories at finite temperature both in the real-time formalism \cite{vanHees:2001ik,VanHees:2001pf} and in the imaginary time formalism \cite{Blaizot:2003an,Berges:2005hc}. It was then extended to the case of scalar theories coupled to fermions \cite{Reinosa:2005pj} and also to abelian gauge theories \cite{Reinosa:2006cm}. Another important extension consisted in including the possibility of non-vanishing field expectation values, of relevance for the determination of the effective potential and thus the study of phase transitions \cite{vanHees:2002bv,Berges:2005hc,Cooper:2005vw,Arrizabalaga:2006hj,Patkos:2008ik,Fejos:2009dm}. The renormalization of models with more complicated global symmetry was studied in \cite{Fejos:2007ec}. From all these studies, a certain consensus has emerged that the renormalization of approximations based on the 2PI effective action seems always possible provided certain extensions of the renormalization procedure are allowed in order to cope with certain approximation artifacts. For instance, in the presence of non-vanishing field expectation values, it is well known that there exist different expressions for the two- and four-point functions, and more generally for higher $n$-point functions. Although equivalent in the exact theory, these various expressions differ within a given approximation and bring their own divergences which can only be absorbed by allowing for apparently more counterterms than usual \cite{Berges:2005hc}. Some of these counterterms are fixed by means of the usual renormalization conditions. The others are fixed by imposing consistency conditions, that is conditions which would be satisfied automatically if no approximation was considered at all \cite{Berges:2005hc,Reinosa:2009tc}. The fact that the consistency conditions do not involve the parameters of the theory is crucial to maintain the number of such parameters to its expected value, despite the larger number of counterterms. In fact, these apparently different counterterms should be viewed as different subseries of the complete perturbative counterterm series and thus should agree when no truncation of the 2PI effective action is considered.\\

In this paper we illustrate partly these issues regarding renormalization by revisiting one of the simplest approximations of the 2PI effective action at non-zero field and finite temperature in the case of a real scalar field, the so-called Hartree approximation. The calculation of the finite temperature effective potential of a real scalar $\varphi^4$ theory in the Hartree approximation and the order of the phase transition that it predicts have been discussed many times in the literature \cite{Espinosa:1992gq,AmelinoCamelia:1992nc,Verschelde:2000ta,Smet:2001un,Cooper:2002qd}. Similar discussions exist for different approximations in the $O(N)$ model with $N\geq 2$, see \cite{Baym:1977qb}. There is a wide consensus on the fact that the order  of the phase transition predicted by this approximation is first order, even though all existing studies are either based on numerical evaluations for particular values of the parameters or involve a high temperature expansion which is not justified for all values of parameters. Part of the originality of this work is that, without relying on a high temperature expansion, it presents a complete analytic confirmation of these results, in the whole parameter space.\\

In section~\ref{sec:2PI}, we briefly recall how the effective potential is computed using the 2PI effective action. The main ingredient is the resolution of a self-consistent equation for the two-point function, the so-called gap equation, which encodes the resummation of particular classes of Feynman diagrams. The Hartree approximation corresponds to the leading order of a systematic approximation scheme for evaluating the 2PI effective action and the corresponding gap equation is then an equation for a self-consistent mass. This equation encodes the resummation of superdaisy diagrams \cite{Dolan:1973qd,AmelinoCamelia:1992nc,Drummond:1997cw} and is discussed at length in section~\ref{sec:gap}. We first recall how renormalization of the self-consistent mass is performed at finite temperature and non-vanishing field expectation value, following a slightly different presentation than the one which is usually found in the literature (although equivalent in practice). We put the emphasis on the fact that not all the divergences of the gap equation lead to divergences of the self-consistent mass. The quadratic divergence of the gap equation is indeed responsible for a divergence of the self-consistent mass, which needs to be absorbed in the renormalization of the bare mass, as usual (sections~\ref{sec:mass} and \ref{sec:triviality}). In contrast, the remaining logarithmic divergence of the gap equation does not lead to a divergence of the self-consistent mass. Instead, it makes this self-consistent mass trivial in the continuum limit, that is equal to the renormalized mass, independently of the value of the temperature or the field expectation value. In this respect, the renormalization of the bare coupling does not appear as a way to absorb any divergence in the self-consistent mass but as a way to avoid triviality (sections~\ref{sec:triviality} and \ref{sec:Landau}). As it is well known, the price to pay for defining a non-trivial scalar theory by means of coupling renormalization is that the theory needs to be regarded as a cut-off theory, with the cut-off $\Lambda$ taken below a certain scale, known as  the Landau scale $\Lambda_{\rm p}$. It is then important to discuss how calculations can be made insensitive to $\Lambda$ in this context. This is discussed in section~\ref{ss:UVity} where we provide a complete analytical discussion of the solutions of the gap equation for any value of $\Lambda$, below and above the Landau scale. Finally, in section~\ref{sec:flow}, we present a new look at the gap equation based on ``evolution'' or ``flow'' equations for the thermal and field dependence of the self-consistent mass: these equations not only provide an efficient way to solve the gap equation but they also shed a new light on its renormalization. Section \ref{sec:eff_pot} is then devoted to the analysis of the effective potential in the Hartree approximation. Its renormalization is described in sections~\ref{sec:geom}  and \ref{sec:ren_eff_pot} with an emphasis on the links with the more general approach presented in \cite{Berges:2005hc} and the distinction between renormalization and consistency conditions. Sections~\ref{sec:extrema} and \ref{sec:temp} discuss analytically the shape of the effective potential as the temperature is lowered from an initial ``high'' temperature $T_\star$ where the theory is chosen to be in the symmetric phase, down to $T=0$. We prove in particular that, depending on the values of the parameters, there is either no transition or a first order phase transition. Section~\ref{sec:concl} is devoted to a discussion concerning the renormalization in the broken symmetry phase and to conclusions.

\vfill

\pagebreak

\section{The 2PI-resummed effective potential}\label{sec:2PI}
In what follows, we consider a real scalar $\varphi^4$ theory in four dimensions at finite temperature, defined by the Euclidean action:
\beq
S[\varphi]\equiv\int_0^{1/T} d\tau\int d^3x\left(\frac{1}{2}(\partial_\tau\varphi)^2+\frac{1}{2}(\nabla\varphi)^2+\frac{m_0^2}{2}\,\varphi^2+\frac{\lambda_0}{4!}\,\varphi^4\right),
\eeq
where the inverse temperature sets the size of the compact interval for temporal integration, and $m_0$ and $\lambda_0$ denote respectively the bare mass and the bare coupling. Notice that, in order that the spectrum be bounded from below, one should restrict to $\lambda_0\geq 0$. In what follows we restrict our attention to the case $\lambda_0>0$.\\

The two-particle-irreducible (2PI) formalism provides a representation of the effective potential $\gamma(\phi)$ corresponding to $S[\varphi]$ in terms of 2PI diagrams. More precisely, it is obtained as the value taken by the 2PI functional:
\beq\label{eq:gen}
\gamma[\phi,G]=\frac{m_0^2}{2}\,\phi^2+\frac{\lambda_0}{4!}\,\phi^4+\frac{1}{2}\int_Q^T\ln G^{-1}+\frac{1}{2}\int_Q^T(Q^2+m_0^2)\,G+\Phi[\phi,G;\lambda_0]
\eeq
at its stationary point $G=\bar G_\phi$, that is $\gamma(\phi)=\gamma[\phi,\bar G_\phi]$ with
\beq\label{eq:stat}
0=\left.\frac{\delta\gamma}{\delta G}\right|_{\phi,\,\bar G_\phi}\,.
\eeq
In Eq.~(\ref{eq:gen}), $\phi$ represents a homogeneous field configuration and $G(i\omega_n,q)$ a function of the Matsubara frequency $\omega_n\equiv2\pi nT$ and the three dimensional momentum $q$. We have also adopted the notation:
\beq\label{eq:notation}
\int_Q^T f(Q)\equiv T\sum_n\int_q\,f(i\omega_n,q)\equiv T\sum_n\int \frac{d^3q}{(2\pi)^3}\,f(i\omega_n,q)\,.
\eeq
Finally, the functional $-\Phi[\phi,G;\lambda_0]$ corresponds to all $0$-leg 2PI diagrams that one can draw in the shifted theory $S[\phi+\varphi]-S[\phi]-(\delta S/\delta\phi)\varphi$ at finite temperature with propagator $G$. This functional cannot be computed exactly. So-called $\Phi$-derivable approximations consist in retaining in $\Phi[\phi,G;\lambda_0]$ only certain classes of diagrams. In this paper we consider the well known Hartree approximation which corresponds to the truncation:
\beq\label{eq:Hartree}
\Phi[\phi,G;\lambda_0]=\frac{\lambda_0}{4}\,\phi^2\int^T_Q G+\frac{\lambda_0}{8}\left(\int_Q^T G\right)^2.
\eeq
According to the above discussion, in order to compute the corresponding effective potential, we first need to determine the stationary propagator $\bar G_\phi$.

\section{Gap equation}\label{sec:gap}
The stationarity condition (\ref{eq:stat}) can be expressed equivalently as
\beq
\bar G^{-1}_{\phi,\,T}(Q)=Q^2+m_0^2+\left.\frac{2\delta\Phi}{\delta G(Q)}\right|_{\bar G_{\phi,\,T}}\,,
\eeq
where $Q^2\equiv\omega_n^2+q^2$. We have used the subscripts $\phi$
and $T$ to stress the fact that the solution $\bar G_{\phi,\,T}$
depends on both the field $\phi$ and the temperature $T$. In what
follows, we shall omit this notation, unless specifically needed. In
the case of the Hartree approximation (\ref{eq:Hartree}), the
propagator takes a very simple form, namely $\bar
G(Q)=1/(Q^2+\bar M^2),$ with the mass $\bar M$ obeying the so-called
gap equation:
\beq\label{eq:gap}
\bar M^2=m_0^2+\frac{\lambda_0}{2}\left[\phi^2+\int_Q^T \bar G\right].
\eeq
We shall explain how to solve this equation later, in Section~\ref{sec:flow}. Before doing so, a fundamental difficulty needs to be bypassed, namely the fact that the gap equation only makes sense within a given ultraviolet regularization. It follows that its solution(s) can depend strongly on the chosen regularization, unless something is done to remove this sensitivity. This is what renormalization is all about in this context.\\

Let us choose for instance a three dimensional, rotation invariant regularization. After performing the Matsubara sum, the sum-integral entering the gap equation reads explicitly:\footnote{Depending on the context, we use the notations $Q<\Lambda$ and $q<\Lambda$ to designate the same three dimensional regularization.}
\beq\label{eq:mat}
\int_{Q<\Lambda}^T \bar G\equiv\int_{q<\Lambda}\frac{1+2n_{\varepsilon_q}}{2\varepsilon_q}\,,
\eeq
where $\varepsilon_q\equiv (q^2+\bar M^2)^{1/2}$ and $n(\varepsilon)\equiv 1/(e^{\beta\varepsilon}-1)$. The term corresponding to $``1"$ in the numerator of Eq.~(\ref{eq:mat}) is particularly sensitive to $\Lambda$. An explicit calculation leads to
\beq
\int_{q<\Lambda}\frac{1}{2\varepsilon_q}=\frac{1}{8\pi^2}\left[\Lambda\sqrt{\Lambda^2+\bar M^2}-\bar M^2{\rm Arcsinh}\left(\frac{\Lambda}{\bar M}\right)\right],\label{eq:vac}
\eeq
which shows that the gap equation (\ref{eq:gap}) possesses terms which depend quadratically and logarithmically on the scale $\Lambda$. What is meant by this is that if we would take $\Lambda$ to infinity, for fixed $m^2_0$, $\lambda_0$, and $\bar M^2$, the right-hand-side of the gap equation would diverge quadratically and logarithmically. However, one should keep in mind that not all these divergences lead to a divergence of the solution of the gap equation.\footnote{It is true that these divergent terms lead to quadratic and logarithmic divergences in the coefficients of the formal perturbative expansion of the solution $\bar M^2$ in powers of $\lambda_0$. However it is not true that they all lead to a divergence of $\bar M^2$, see the subsequent discussion.} The purpose of the remainder of this section is to clarify the connection between the sensitivity of the gap equation to the scale $\Lambda$ and the sensitivity of its solution(s) to this very same scale, and to explain how the later can be removed, or at least substantially reduced.

\subsection{Quadratic divergence}\label{sec:mass}
Let us first prove that the quadratic divergence of the gap equation (\ref{eq:gap}) leads to a divergence of its solution $\bar M^2$. We first prove that there is a unique solution for large enough $\Lambda$ and then that this solution diverges with increasing $\Lambda$. To this purpose, it is convenient to write the gap equation as $0=f_\Lambda(\bar M^2)$ with
\beq\label{eq:deff}
f_\Lambda(M^2)\equiv-M^2+m_0^2+\frac{\lambda_0}{2}\left[\phi^2+\int_{Q<\Lambda}^T G\right]
\eeq
and $G\equiv 1/(Q^2+M^2)$, and study the positive\footnote{We shall only be concerned with positive ($\bar M^2\geq 0$) solutions of the gap equation. For a discussion of negative solutions of the ``explicit'' form of the gap equation at zero temperature, that is the gap equation where the integrals have been performed explicitly for $M^2>0$ and then extended to any value of $M^2$, see \cite{Nunes:1993bk}.} zeros of $f_\Lambda(M^2)$. The derivative of this function with respect to $M^2$ reads
\beq\label{eq:fprime}
f'_\Lambda(M^2)=-1-\frac{\lambda_0}{2}\int_{Q<\Lambda}^T G^2\,,
\eeq
which is always strictly negative because $\lambda_0>0$. It follows that $f_\Lambda(M^2)$ decreases strictly from
\beq
f_\Lambda(0)=m_0^2+\frac{\lambda_0}{2}\left[\phi^2+\int_{Q<\Lambda}^T\frac{1}{Q^2}\right]
\eeq
to $f_\Lambda(\infty)=-\infty$ (the regularized tadpole integral in
Eq.~(\ref{eq:deff}) is suppressed for $M^2\gg\Lambda^2,\,T^2$). Then,
the existence of a solution of the gap equation depends on the sign of
$f_\Lambda(0)$. Even though the parameter $m_0^2$ could be negative,
the quadratic and positive divergence in $f_\Lambda(0)$, given 
explicitly in Eq.~(\ref{eq:vac}), implies that there exists a
value of $\Lambda$ above which $f_\Lambda(0)\geq 0$ and the gap
equation admits a unique solution $\bar M^2$. For the same reason,
given a mass $\mu$, there exists a value of $\Lambda$ above which
$f_\Lambda(\mu^2)\geq 0$ and thus $\bar M^2\geq \mu^2$. This shows
that the solution $\bar M^2$ diverges as $\Lambda\rightarrow\infty$,
for fixed $m^2_0$ and $\lambda_0$, as announced.

\subsection{Renormalization of the mass -- Triviality}\label{sec:triviality}
Since the quadratic divergence of the gap equation depends neither on $\phi$ nor on $T$, it can be absorbed by adjusting the divergent part of $m^2_0$. On the other hand, the finite part of $m^2_0$ can be used to impose a condition at a given value of $\phi$ and a given value of $T$. We choose $\phi=0$, $T=T_\star$ and impose the renormalization condition: 
\beq\label{eq:rencondm}
\bar M^2_{\phi=0,\,T_\star}=m^2_\star\,.
\eeq
The parameter $m^2_\star$ is positive by construction, because it is a solution of the gap equation at $\phi=0$ and $T=T_\star.$ We shall choose it strictly positive in what follows. The renormalization condition (\ref{eq:rencondm}) is quite natural when studying how the system evolves as the temperature $T$ is decreased from a ``high'' temperature $T_\star$ where the system is required to be in the symmetric phase, see Section~\ref{sec:eff_pot}. It can be rewritten as a choice of the bare mass, namely:
\beq\label{eq:m0}
m_0^2=m_\star^2-\frac{\lambda_0}{2}\int_{Q<\Lambda}^{T_\star} G_\star\,,
\eeq
with $G_\star\equiv 1/(Q^2+m^2_\star)$. With this choice, the gap equation reads $0=\tilde f_\Lambda(\bar M^2)$, with
\beq\label{eq:gapnm}
\tilde f_\Lambda(M^2)\equiv-M^2+m_\star^2+\frac{\lambda_0}{2}\left[\phi^2+\int_{Q<\Lambda}^T G-\int_{Q<\Lambda}^{T_\star}G_\star\right].
\eeq
The dependence of the gap equation on $\Lambda$ has been changed by the renormalization procedure and we
need to reconsider its possible solutions and their dependence on $\Lambda$. As before, we do so by discussing the zeros of the function $\tilde f_\Lambda(M^2)$.\\

Notice first that $\tilde f'_\Lambda(M^2)=f'_\Lambda(M^2)$ is strictly negative in view of Eq.~(\ref{eq:fprime}). It follows that the function $\tilde f_\Lambda(M^2)$ decreases strictly from
\beq
\tilde f_\Lambda(0)=m_\star^2+\frac{\lambda_0}{2}\left[\phi^2+\int_{Q<\Lambda}^T\frac{1}{Q^2}-\int_{Q<\Lambda}^{T_\star}G_\star\right]
\eeq
to $\tilde f_\Lambda(\infty)=-\infty$ (the regularized tadpole
integral in Eq.~(\ref{eq:gapnm}) is suppressed for
$M^2\gg\Lambda^2,\,T^2$). Using Eqs.~(\ref{eq:mat}) and
(\ref{eq:vac}), it is easily checked that $\tilde f_\Lambda(0)$
diverges logarithmically like $c\,m_\star^2\ln\Lambda,$ with
$c>0$. Then, there exists a value of $\Lambda$ above which $\tilde
f_\Lambda(0)\geq 0$ and the gap equation admits a unique solution
$\bar M^2$. For the same reason, for any $\Delta m^2>0$, $\tilde
f_\Lambda(m_\star^2\pm\Delta m^2)$ diverges logarithmically like
$\mp\,c\,\Delta m^2\ln\Lambda$. Then, there exists a value of
$\Lambda$ above which $\tilde f_\Lambda(m^2_\star-\Delta m^2)\geq 0$
and $\tilde f_\Lambda(m^2_\star+\Delta m^2)\leq 0$, and thus $|\bar
M^2-m^2_\star|\leq\Delta m^2$. This proves that $\bar M^2\rightarrow
m^2_\star$ as $\Lambda\rightarrow\infty$, for fixed $\lambda_0$. Using
the gap equation $0=\tilde f_\Lambda(\bar M^2)$ as well as
Eqs.~(\ref{eq:mat}) and (\ref{eq:vac}), it is possible to determine
precisely how this limit is approached. We obtain
\beq\label{eq:approach}
\bar M^2-m^2_\star\sim\frac{8\pi^2}{\ln\Lambda}\left[\phi^2+\int_{q}\frac{\delta_\star n_{\varepsilon^\star_q}}{\varepsilon^\star_q}\right]\,,
\eeq
where $\varepsilon^\star_q\equiv (q^2+m^2_\star)^{1/2}$ and $\delta_\star n_\varepsilon\equiv n_\varepsilon-n^\star_\varepsilon$, with $n^\star_\varepsilon$ the thermal factor at temperature $T_\star$. Thus, unlike what happened with the quadratic divergence, the logarithmic divergence of the gap equation (the one that remains after mass renormalization) does not lead to a divergence of the solution $\bar M^2$.\footnote{The absence of divergence in the solution of the mass renormalized gap equation means that the resummation of superdaisy diagrams encoded in this equation formally resums the perturbative logarithmic divergences associated with each diagram of this series into a convergent contribution. That perturbative divergences can be resummed to non-divergent expressions was observed in \cite{Ghinculov:1997tn,Fejos:2009dm}.} Rather, if we insist in taking the limit $\Lambda\rightarrow\infty$ for fixed $\lambda_0$, the solution of the gap equation converges to the renormalized mass $m^2_\star$ for any value of the field $\phi$ and the temperature $T$. This illustrates the triviality of $\varphi^4$ theory \cite{Drummond:1997cw}, at least in the particular approximation considered here.  Equation (\ref{eq:approach}) shows that the trivial limit is approached rather slowly (logarithmically).

\subsection{Renormalization of the coupling -- Landau pole}\label{sec:Landau}
The previous analysis has shown that, for the particular choice of mass renormalization we have considered, the triviality of $\varphi^4$ theory is related to the presence of a logarithmically divergent term in the gap equation. In order to define a non-trivial theory, one can absorb this divergence in a redefinition of $\lambda_0$. Using the formula (see Appendix~\ref{app:id}):
\beq\label{eq:id}
\int_{Q<\Lambda}^T \bar G=\int_{Q<\Lambda}^{T_\star}\bar G+\int_{q<\Lambda}\frac{\delta_\star n_{\varepsilon_q}}{\varepsilon_q}\,,
\eeq
as well as the identity:
\beq\label{eq:id3}
\bar G-G_\star=-(\bar M^2-m^2_\star)\,G_\star\,\bar G=-(\bar M^2-m^2_\star)\,G_\star^2+(\bar M^2-m^2_\star)^2\,G_\star^2\,\bar G\,,
\eeq
it is clear that the logarithmic divergence in Eq.~(\ref{eq:gapnm}) is entirely accounted for by the term $-(\bar M^2-m^2_\star)\int_{Q<\Lambda}^{T_\star}G_\star^2$. Isolating this contribution and defining an effective coupling $\lambda_\star$ \cite{Coleman:1974jh} such that
\beq\label{eq:effl}
\frac{1}{\lambda_\star}\equiv\frac{1}{\lambda_0}+\frac{1}{2}\int_{Q<\Lambda}^{T_\star} G_\star^2\,,
\eeq
the gap equation $0=\tilde f_\Lambda(\bar M^2)$ can be rewritten as $0=g_\Lambda(\bar M^2)$ with
\beq\label{eq:gapr}
g_\Lambda(M^2) & \equiv & -M^2+m^2_\star+\frac{\lambda_\star}{2}\left[\phi^2+\int_{Q<\Lambda}^T G-\int_{Q<\Lambda}^{T_\star}G_\star+(M^2-m^2_\star)\int_{Q<\Lambda}^{T_\star}G_\star^2\right]\nonumber\\
& = & -M^2+m^2_\star+\frac{\lambda_\star}{2}\left[\phi^2+\int_{q<\Lambda}\frac{\delta_\star n_{\varepsilon_q}}{\varepsilon_q}+(M^2-m^2_\star)^2\int^{T_\star}_{Q<\Lambda} G_\star^2\,G\right],
\eeq
where, in the second line, we have made use of Eqs.~(\ref{eq:id}) and (\ref{eq:id3}) to obtain an explicitly convergent expression that we will use later. If we insist in keeping the bare coupling $\lambda_0$ fixed and positive, then, according to Eq.~(\ref{eq:effl}), the effective coupling $\lambda_\star$ goes to zero as $\Lambda\rightarrow\infty$ and we recover the trivial result with $\bar M^2-m^2_\star\sim\lambda_\star\,[\phi^2+\int_q \delta_\star n_{\varepsilon^\star_q}/\varepsilon^\star_q]/2$, in agreement with Eq.~(\ref{eq:approach}). In contrast, fixing the value of $\lambda_\star$ allows to avoid triviality. This corresponds to the following redefinition of $\lambda_0$ ($\lambda_0$ is then $\phi$- and $T$-independent as it should):
\beq\label{eq:l0}
\frac{1}{\lambda_0}=\frac{1}{\lambda_\star}-\frac{1}{2}\int_{Q<\Lambda}^{T_\star} G_\star^2\,.
\eeq
However, it appears that in order to maintain $\lambda_0>0$, and in turn a meaningful microscopic theory, one needs to restrict the cut-off $\Lambda$ below a certain scale known as the Landau scale or Landau pole $\Lambda_{\rm p}$, defined by
\beq\label{eq:Landau}
0=\frac{1}{\lambda_\star}-\frac{1}{2}\int_{Q<\Lambda_{\rm p}}^{T_\star}G_\star^2\,.
\eeq
In other words, the non-trivial Hartree approximation has a meaning only if
it is considered as describing an effective theory. Notice that, from
Eq.~(\ref{eq:Landau}), it follows that $\lambda_\star>0$ (negative values of the  renormalized coupling could be possible with other renormalization schemes without violating the requirement $\lambda_0>0$).  More explicitly, using Eq.~(\ref{eq:id2}), we obtain
\beq
\frac{1}{\lambda_\star}=\frac{1}{16\pi^2}\left[{\rm Arcsinh}\left(\frac{\Lambda_{\rm p}}{m_\star}\right)-\frac{\Lambda_{\rm p}}{\sqrt{\Lambda^2_{\rm p}+m^2_\star}}\right]+\frac{1}{2}\int_{q<\Lambda_{\rm p}}\frac{n^\star_{\varepsilon^\star_q}-\varepsilon^\star_q{n^\star}'_{\varepsilon^\star_q}}{2{\varepsilon^\star_q}^3}\,.
\eeq
In the limit $\lambda_\star\rightarrow 0^+$, $\Lambda_{\rm p}\rightarrow\infty$ with $\Lambda^2_{\rm p}\sim\mu^2_\star\,e^{32\pi^2/\lambda_\star}$ where $\mu_\star$ depends on $m_\star$ and $T_\star$:
\beq
\mu^2_\star=\frac{m^2_\star}{4}\exp\left(2-16\pi^2\int_q\frac{n^\star_{\varepsilon^\star_q}-\varepsilon^\star_q{n^\star}'_{\varepsilon^\star_q}}{2{\varepsilon^\star_q}^3}\right).
\eeq

\subsection{Ultraviolet sensitivity \label{ss:UVity}}
Because, in the Hartree approximation at least, a non-trivial $\varphi^4$ theory is only to be considered as a
cut-off theory, we need to wonder how calculations done within such a
theory can be made almost independent of the cut-off $\Lambda$. The
main idea is that, if one considers the regime $\Lambda_{\rm p}\gg
T_\star,\,m_\star,\,\phi,\,T$, one can choose the cut-off $\Lambda$
such that both requirements $\Lambda_{\rm p}>\Lambda$ and $\Lambda\gg
T_\star,\,m_\star,\,\phi,\,T$ are met. Then, because all divergences of the
gap equation have been absorbed in the redefinition of the bare
parameters, we expect that its solution $\bar M^2$ be almost cut-off
independent, up to terms of order
$T_\star/\Lambda\,,m_\star/\Lambda\,,\phi/\Lambda\,,T/\Lambda\ll
1$. The Hartree approximation offers the possibility to illustrate this issue, for one can study its solutions as a function of $\Lambda$, in particular as $\Lambda\rightarrow\infty$, and thus assess under which conditions these solutions can be considered practically insensitive to $\Lambda$ for $\Lambda<\Lambda_{\rm p}$.\\

\begin{figure}[htbp]
\begin{center}
\includegraphics[keepaspectratio,width=0.7\textwidth,angle=0]{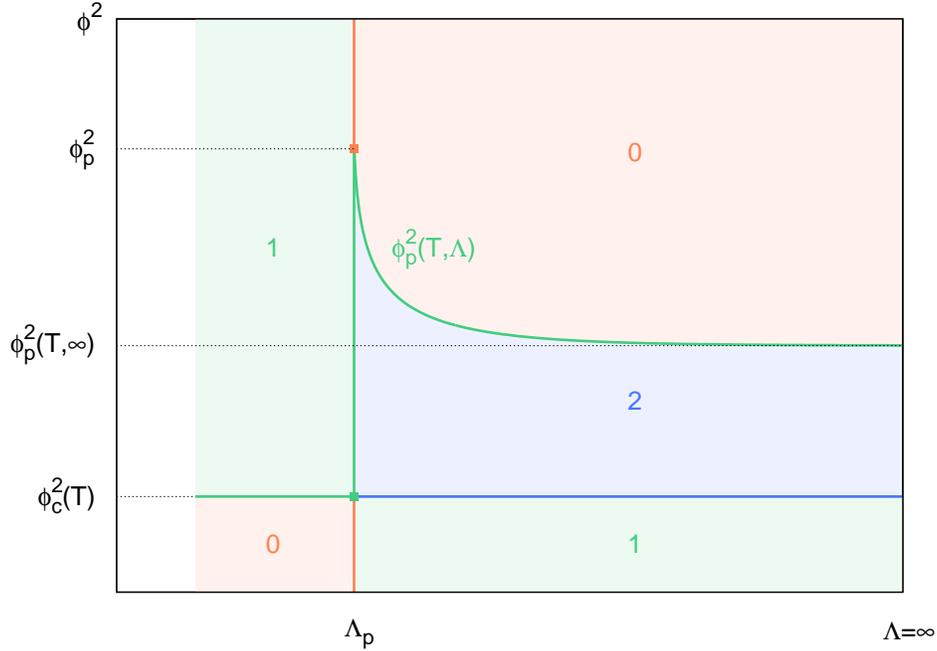}
\caption{Number of solutions of the gap equation $0=g_\Lambda(\bar M^2)$ 
in the $(\Lambda,\phi^2)$-plane in the regime of interest,
that is when $\Lambda,\Lambda_{\rm p}\gg T_\star,m_\star,\phi,T.$
Orange areas correspond to no solution, green areas
correspond to one solution, and the blue area corresponds to two
solutions, as indicated also by the labels. In the regime of interest $\phi^2_{\rm p}\sim\Lambda_{\rm p}^2/(8\pi^2)$ and $\phi^2_{\rm p}(T,\infty)\sim\Lambda_{\rm p}^2/(4\pi^2 e^2)$ are both high scales.\label{fig:sols}}
\end{center}
\end{figure}

The analysis of
the solutions of the gap equation $0=g_\Lambda(\bar M^2)$ as one
varies the cut-off $\Lambda$ is given in Appendix~\ref{app:sol} where, in order to
simplify the discussion, we restrict our analysis to the ``regime of
interest'' that is a regime where both $\Lambda$ and $\Lambda_{\rm p}$
are much larger than all the other scales $T_\star$, $m_\star$, $\phi$,
and $T$. The result of this analysis is that the number of solutions of the gap equation depends on the values of $\Lambda$ and $\phi^2,$ as we represent in Fig.~\ref{fig:sols} and explain in what follows:\\

1. For $\Lambda<\Lambda_{\rm p}$, there is a ``critical'' value
$\phi^2_{\rm c}(T,\Lambda)$ of $\phi^2$ such that the gap equation
admits a unique solution if $\phi^2\geq\phi^2_{\rm c}(T,\Lambda)$ and
no solution if $\phi^2<\phi^2_{\rm c}(T,\Lambda)$, see
Fig.~\ref{fig:sols}. Notice that $\phi^2_{\rm c}(T,\Lambda)$ is not
necessarily positive. In practice it is useful to know when it could
become strictly positive, signaling the fact that the gap equation has no (positive) solutions if $0\leq\phi^2<\phi^2_{\rm c}(T,\Lambda)$. To this purpose, one writes, see Appendix~\ref{app:sol}:
\beq
\phi^2_{\rm c}(T,\Lambda)=-\frac{2\,C_\star}{\lambda_\star}-\int_{q<\Lambda}\frac{n_q}{q}\,,
\eeq
with
\beq\label{eq:Cstar}
C_\star\equiv m^2_\star+\frac{\lambda_\star}{2}\left[-\int_{q<\Lambda}\frac{n^\star_q}{q}+m^4_\star \int^{T_\star}_{Q<\Lambda} \frac{G_\star^2}{Q^2}\right].
\eeq
Clearly, $\phi^2_{\rm c}(T,\Lambda)$ is strictly positive if the parameters $\Lambda$, $T_\star$, $m_\star$, and $\lambda_\star$ are such that $C_\star<0$ and the temperature $T$ is strictly below a certain ``critical'' temperature $T_{\rm c}$ defined by
\beq\label{eq:Tc}
\int_{q<\Lambda}\frac{n^{\rm c}_q}{q}\equiv-\frac{2\,C_\star}{\lambda_\star}\,.
\eeq
Using Eq.~(\ref{eq:Cstar}), it is easy to check that $\phi^2_{\rm c}(T_\star,\Lambda)$ is strictly negative. It follows that $T_{\rm c}<T_\star$. Moreover, in the regime of interest, one can neglect the dependence of $C_\star$ and of the thermal integrals with respect to $\Lambda$. The critical value $\phi^2_{\rm c}(T,\Lambda)$ is then independent of $\Lambda$ (this is the reason why it is represented by an horizontal line in Fig.~\ref{fig:sols}) and one has $\phi^2_{\rm c}(T)=(T^2_{\rm c}-T^2)/12$ with $T^2_{\rm c}=-24C_\star/\lambda_\star$.\\

2. For $\Lambda=\Lambda_{\rm p}$, a new ``relevant'' scale appears:
\beq
\phi^2_{\rm p}\equiv\int_{Q<\Lambda_{\rm p}}^{T_\star}G_\star>0\,,
\eeq
see Appendix~\ref{app:sol}. It is such that $\phi^2_{\rm p}>\phi^2_{\rm c}(T)$ and the gap equation admits no solution if $\phi^2\geq\phi^2_{\rm p}$, one solution if $\phi^2_{\rm p}>\phi^2\geq\phi^2_{\rm c}(T)$, and again no solution if $\phi^2<\phi^2_{\rm c}(T)$, see Fig.~\ref{fig:sols}. Notice that, in the regime of interest, $\phi^2_{\rm p}\sim\Lambda^2_{\rm p}/(8\pi^2)$ is a high scale.\\ 

3. Similarly, for $\Lambda>\Lambda_{\rm p}$ a new relevant scale $\phi^2_{\rm p}(T,\Lambda)>\phi^2_{\rm c}(T)$ enters the discussion. The gap equation admits no solution if $\phi^2>\phi^2_{\rm p}(T,\Lambda)$ one solution if $\phi^2=\phi^2_{\rm p}(T,\Lambda)$, two solutions if $\phi^2_{\rm p}(T,\Lambda)>\phi^2\geq\phi^2_{\rm c}(T)$ and one solution if $\phi^2<\phi^2_{\rm c}(T)$, see Fig.~\ref{fig:sols}. The value of $\phi^2_{\rm p}(T,\Lambda)$ as $\Lambda\rightarrow\Lambda_{\rm p}^+$ is nothing but $\phi^2_{\rm p}$. Moreover, in the regime of interest, one can show that $\phi^2_{\rm p}(T,\Lambda)$ decreases with $\Lambda$, from $\phi^2_{\rm p}(T,\Lambda^+_{\rm p})=\phi^2_{\rm p}$ to $\phi^2_{\rm p}(T,\infty)\sim\Lambda^2_{\rm p}/(4\pi^2e^2)$, another high scale.\\

\begin{figure}[htbp]
\begin{center}
\includegraphics[keepaspectratio,width=0.495\textwidth,angle=0]{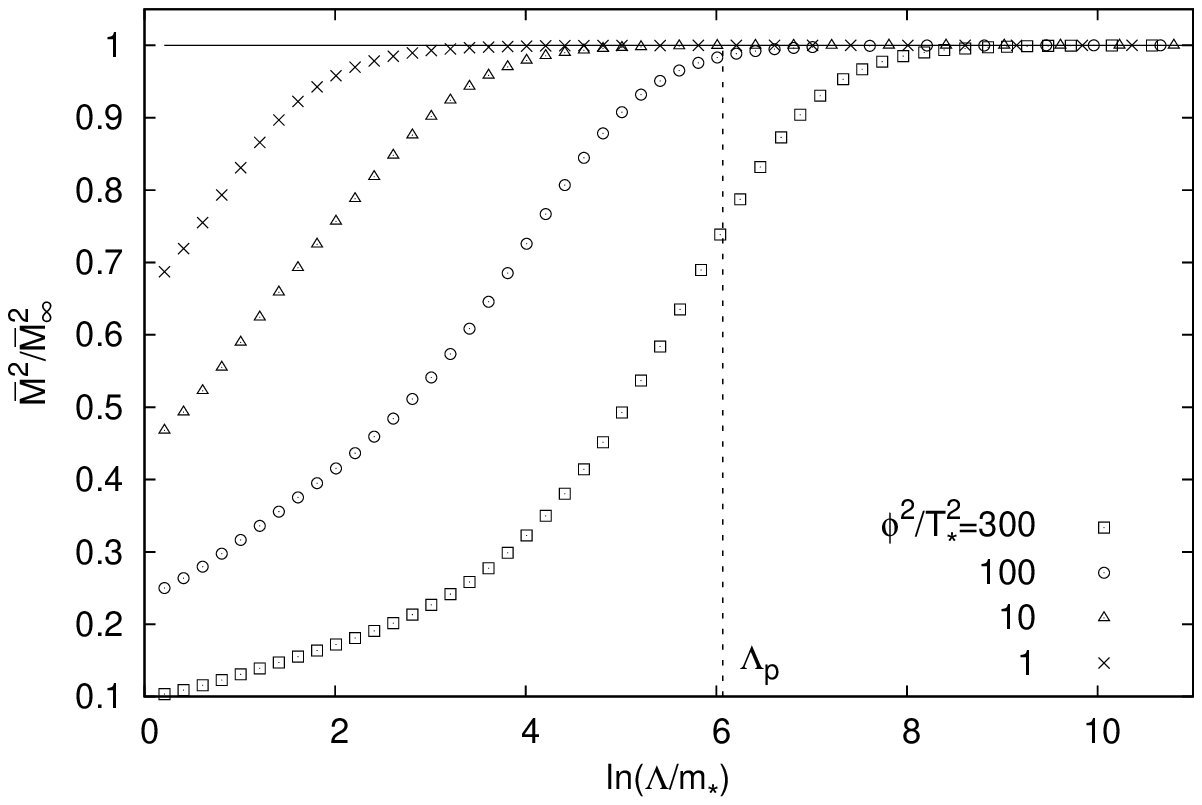}       
\includegraphics[keepaspectratio,width=0.495\textwidth,angle=0]{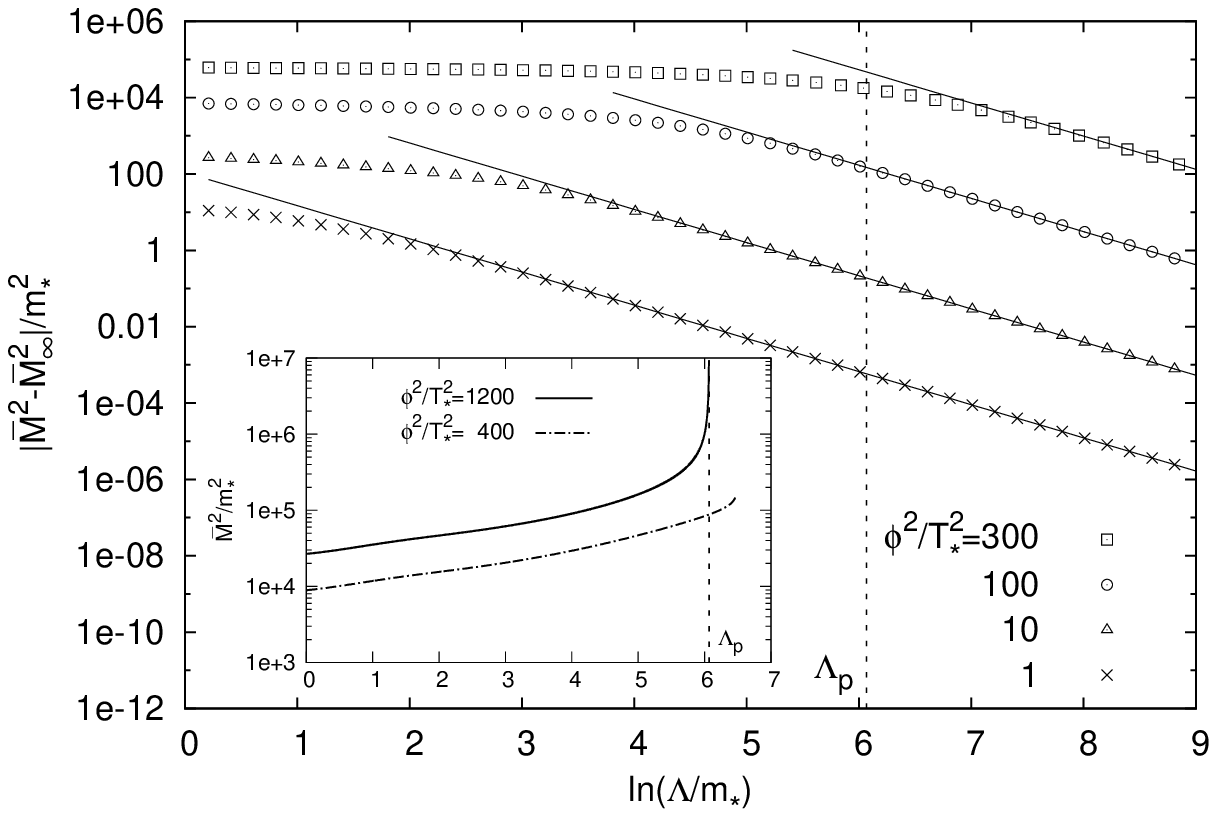}
\caption{ Convergence of the physical solution of the renormalized and
$\Lambda$-dependent gap equation at different values of the field for
$m_\star^2/T_\star^2=0.5,$ $\lambda_\star=20$ and $T/T_\star=0.012$.  The dahed vertical line, common to both plots, indicates the location of the Landau scale. The left panel shows the dependence of the rescaled physical solution with respect to the cut-off $\Lambda$. The right panel illustrates the validity of the asymptotic formula (\ref{eq:convergence}) represented by the solid lines. The inset shows the behavior of the
solution for two values of the field which lie outside of the
convergence window (see text for details).
\label{fig:conv}}
\end{center}
\end{figure}

The previous analysis shows in particular that the unique solution of the gap equation which exists for $\Lambda<\Lambda_{\rm p}$ and $\phi^2\geq\phi^2_{\rm c}(T)$ and which we shall call the {\it physical solution}, behaves rather differently, depending on the value of $\phi^2$, as one increases the value of $\Lambda$. If $\phi^2\geq\phi^2_{\rm p}$, as we show in Appendix \ref{app:sol}, the physical solution diverges as $\Lambda\rightarrow\Lambda_{\rm p}^-$. In contrast, if $\phi^2_{\rm p}>\phi^2\geq \phi^2_{\rm c}(T)$, nothing of this kind occurs: the physical solution can be extended beyond the Landau scale and a second solution, which we call the {\it unphysical solution}, appears. If $\phi^2_{\rm p}>\phi^2>\phi^2_{\rm p}(\infty)$, there is value of $\Lambda$ at which the two solutions merge into one solution and then cease to exist (without diverging). On the contrary, in the {\it convergence window} $\phi^2_{\rm p}(\infty)\geq\phi^2\geq\phi^2_{\rm c}(T)$ the two solutions can be extended to $\Lambda\rightarrow\infty$ and one can then assess how rapidly they converge to their limiting values $\bar M^2_\infty$. To do so, we take a derivative of the gap equation $0=g_\Lambda(\bar M^2)$ with respect to $\Lambda$. We obtain
\beq
\frac{\partial\bar M^2}{\partial\Lambda}=-\frac{1}{g'_\Lambda(\bar M^2)}\frac{\partial g_\Lambda}{\partial\Lambda}\,,
\eeq
where $\partial g_\Lambda/\partial\Lambda$ corresponds to the explicit dependence of $g_\Lambda$ with respect to $\Lambda$. An implicit dependence is also present, through $\bar M^2$, which explains the appearance of $g'_\Lambda(\bar M^2)$ in the previous formula. We are allowed to divide by $g'_\Lambda(\bar M^2)$ because, as long as $\phi^2<\phi^2_{\rm p}(\infty)$, one has $g'_\Lambda(\bar M^2)\neq 0$, see Appendix~\ref{app:sol}. Moreover, this remain true in the limit $\Lambda\rightarrow\infty$. It follows that
\beq
\frac{\partial\bar M^2}{\partial\Lambda}\sim-\frac{1}{g'_\infty(\bar M^2_\infty)}\left.\frac{\partial g_\Lambda}{\partial\Lambda}\right|_{\bar M^2=\bar M^2_\infty}\,.
\eeq
Then, a simple calculation using Eqs.~(\ref{eq:id1}), (\ref{eq:id2}) and (\ref{eq:un}) leads to
\beq
\frac{\partial\bar M^2}{\partial\Lambda}\sim-\frac{3}{64\pi^2 g'_\infty(\bar M^2_\infty)}\frac{(\bar M^2_\infty-m_\star^2)^2}{\Lambda^3}\,,
\eeq
or, in other words:
\beq\label{eq:convergence}
\left|\frac{\bar M^2}{m_\star^2}-\frac{\bar M^2_\infty}{m_\star^2}\right|\sim \frac{3\lambda_\star}{128\pi^2|g'_\infty(\bar M^2_\infty)|}\left(1-\frac{\bar M^2_\infty}{m_\star^2}\right)^2\frac{m_\star^2}{\Lambda^2}\,,
\eeq
where we have expressed all the masses in units of the renormalized
mass $m_\star$. Although we did not make it explicit, the limiting
value $\bar M^2_\infty$ depends on $T$ and $\phi$. In the case of the physical solution $\bar M^2_\infty$ increases
strictly from $0$ at $\phi^2=\phi^2_{\rm c}(T)$ to a very large value
$\bar M^2_{\rm p}(\infty)\sim 4\Lambda^2_{\rm p}/e^2$ at
$\phi^2=\phi^2_{\rm p}(\infty)\sim\Lambda^2_{\rm p}/(4\pi^2e^2)$, see
Appendix~\ref{app:sol}. Thus, in the regime of interest, such as in
particular $\phi\ll\Lambda_{\rm p}$, $\bar M^2_\infty/m^2_\star$ is of
the order $1$ or smaller and the scale multiplying the convergence
factor $1/\Lambda^2$ in Eq.~(\ref{eq:convergence}) is a low scale of
the order of $m^2_\star\ll\Lambda_{\rm p}^2$ or smaller (the factor
$1/g'_\infty(\bar M^2_\infty)$ can only improve the convegence, for
it goes to zero when $\bar M^2_\infty\rightarrow 0$). Then, the
insensitivity of the physical solution to the scale $\Lambda$ is expected to be observed
already for values of $\Lambda$ below the Landau scale 
$\Lambda_{\rm p}$. This expectation is confirmed by the numerical results presented in Fig.~\ref{fig:conv} which shows the convergence of the physical solution as $\Lambda\rightarrow\infty$ for increasing values of $\phi$ : for small values of the field the physical solution has already converged for values
of $\Lambda$ well below the Landau pole, while for very large values
of $\phi$ the convergence, if it happens, occurs for values of $\Lambda$ above the Landau pole. This is a clear illustration of why, although there exists a
Landau pole $\Lambda_{\rm p}$ which forces us to choose the cut-off $\Lambda<\Lambda_{\rm p}$, one can
define, in some region of the parameter space, a theory whose
results are almost insensitive to the cut-off $\Lambda$.\\ 

\subsection{Flow approach to the gap equation}\label{sec:flow}
To conclude this section, we present an elegant way to solve the gap equation. This approach is also interesting on formal grounds because it gives a new perspective on renormalization. It could also be used for higher order approximations.\\

Let us assume that the solution of the gap equation is known for some $\phi$ and $T$, and let us ask how this solution changes as we vary $\phi$ or $T$. The change in the solution is governed by ``evolution'' or ``flow'' equations that we now derive. Let us first express the infinitesimal change of $\bar M^2$ with respect to $\phi^2$. From the original bare gap equation (\ref{eq:gap}) and using the fact that the bare parameters $m^2_0$ and $\lambda_0$ are $\phi$-independent, we obtain
\beq
\frac{\partial \bar M^2}{\partial \phi^2}=\frac{\lambda_0}{2}\left[1+\int_{Q<\Lambda}^T\frac{\partial\bar G}{\partial\phi^2}\right]=\frac{\lambda_0}{2}-\frac{\lambda_0}{2}\int_{Q<\Lambda}^T\bar G^2\,\frac{\partial \bar M^2}{\partial\phi^2}\,,
\eeq
where we used $\delta\bar G=-\bar G^2\,\delta\bar G^{-1}$. Collecting
the derivatives of $\bar M^2$ on the same side of the equation, this can be rewritten as
\beq\label{eq:flowMphi}
\frac{\partial \bar M^2}{\partial \phi^2}=\frac{\bar V}{2}\,,
\eeq
with the function $\bar V$ defined by
\beq\label{eq:BS}
\frac{1}{\bar V}=\frac{1}{\lambda_0}+\frac{1}{2}\int_{Q<\Lambda}^T\bar G^2\,.
\eeq
Similar manipulations are possible to express the infinitesimal change of $\bar M^2$ with respect to $T$. Using the fact that the bare parameters are $T$-independent, one writes
\beq
\frac{\partial \bar M^2}{\partial T}=\frac{\lambda_0}{2}\left(\frac{\partial}{\partial T}\int_{Q<\Lambda}^T\right)\bar G-\frac{\lambda_0}{2}\int_{Q<\Lambda}^T\bar G^2\,\frac{\partial \bar M^2}{\partial T}\,,
\eeq
where the notation $(\partial/\partial T\int_{Q<\Lambda}^T) \bar G$
defined in Appendix~\ref{app:id} refers to a derivative with
respect to the explicit thermal dependence, not the implicit one,
hidden in $\bar G.$ \rt{, see Appendix~\ref{app:id}.} As before, this can be rewritten as:
\beq\label{eq:TflowM}
\frac{\partial \bar M^2}{\partial T}=\frac{\bar V}{2}\left(\frac{\partial}{\partial T}\int_{Q<\Lambda}^T\right)\bar G\,.
\eeq
A similar strategy can be used to express the infinitesimal variations of $\bar V$ itself. We obtain:
\beq
\frac{\partial\bar V}{\partial \phi^2} & = & \frac{\bar V^3}{2}\int_{Q<\Lambda}^T\bar G^3\,,\label{eq:flowVphi}\\
\frac{\partial\bar V}{\partial T} & = & -\frac{\bar V^2}{2}\left(\frac{\partial}{\partial T}\int_{Q<\Lambda}^T\right)\bar G^2+\frac{\bar V^3}{2}\left(\int_{Q<\Lambda}^T\bar G^3\right)\left(\frac{\partial}{\partial T}\int_{Q<\Lambda}^T\right)\bar G\label{eq:TflowV}\,.
\eeq
The interesting feature of Eqs.~(\ref{eq:flowMphi}), (\ref{eq:TflowM}), (\ref{eq:flowVphi}) and (\ref{eq:TflowV}) is that they are explicitly finite, without the need to impose any renormalization condition. They can be used as an alternative tool to solve the gap equation. The only information that remains to be specified are the initial conditions for $\bar M^2$ and $\bar V$, for instance at $\phi^2=0$ and $T=T_\star$. The initial condition for $\bar M^2$ reads $\bar M^2_{\phi=0,\,T_\star}=m^2_\star$ which is nothing but the renormalization condition (\ref{eq:rencondm}). As for $\bar V$, its initial condition reads:
\beq\label{eq:rencondV}
\bar V_{\phi=0,T_\star}=\lambda_\star\,,
\eeq
and it is easily checked that it can be reinterpreted as the the renormalization condition corresponding to the choice of $\lambda_0$ in Eq.~(\ref{eq:l0}). Thus, specifying the  initial conditions in the flow approach corresponds to imposing renormalization conditions in the standard approach\footnote{This is similar in spirit to the approach followed in \cite{Blaizot:2010zx}.}. As already discussed, $m^2_\star$ and $\lambda_\star$ are taken strictly positive.\\

In practice, it is more convenient to solve the evolution equations (\ref{eq:flowMphi}) and (\ref{eq:TflowM}) coupled to the explicit and renormalized expression for $\bar V$ which we obtain from combining Eqs.~(\ref{eq:BS}) and (\ref{eq:l0}):
\beq\label{eq:BSren}
\frac{1}{\bar V}=\frac{1}{\lambda_\star}+\frac{1}{2}\left[\int_{Q<\Lambda}^T\bar G^2-\int_{Q<\Lambda}^{T_\star} G_\star^2\right]\,.
\eeq
Because $\phi^2_{\rm c}(T_\star)<0$, we first solve the equation (\ref{eq:flowMphi}) at $T=T_\star$ from the initial condition $\bar M^2_{\phi=0,\,T_\star}=m_\star^2>0$ and obtain the function $\bar M^2_{\phi,\,T_\star}$. For each value of $\phi$, the solution $\bar M^2_{\phi,\,T_\star}$ is then used as the initial condition for the flow equation (\ref{eq:TflowM}), which we solve from $T_\star$ for decreasing values of $T$. We obtain in this way the function $\bar M^2_{\phi,\,T}$ for any value of $\phi$ and $T$ (with $\phi^2\geq\phi^2_{\rm c}(T)$). This method will be used in the next section where we evaluate the effective potential. Notice finally that $\bar V=-\lambda_\star/g'_\Lambda(\bar M^2)$ where
$g_\Lambda(M^2)$ is the function introduced in Eq.~(\ref{eq:gapr}) to
discuss the solutions of the renormalized gap equation (see
Eq.~(\ref{eq:deux}) for its derivative). We are only interested in the physical solution to the gap equation. Irrespectively of the value of $\Lambda$, this solution is such that $g'_\Lambda(\bar M^2)<0$, see Appendix~\ref{app:sol}, and thus $\bar V>0$. From Eqs.~(\ref{eq:flowMphi}) and (\ref{eq:flowVphi}), it follows then that $\bar M^2$ and $\bar V$ increase with $\phi^2$. In particular at $T=T_\star$ and for any $\phi^2$, one has $\bar M^2\geq m^2_\star$ and $\bar V\geq\lambda_\star$. We will make use of this remark later.

\section{Effective potential}\label{sec:eff_pot}
Once the mass $\bar M$ and the propagator $\bar G$ have been
determined, one can evaluate numerically the renormalized version of the
  effective potential to be given in Section \ref{sec:ren_eff_pot} and study its change of shape as the temperature
$T$ is lowered from the ``high'' temperature $T_\star$, possibly
triggering a change of phase. In this section we show that a large
part of the analysis can be done analytically. We prove in particular
that in the Hartree approximation, for those values of the parameters such that there is a phase transition,
the later cannot be of second order. To this purpose, we use information on
the geometry of the effective potential, such as its curvature, as
well as information on the location of its possible extrema, as given
by the field equation. Notice that our analysis will be performed in the presence of a cut-off $\Lambda$ below the Landau scale $\Lambda_{\rm p}$. After renormalization of the effective potential, see below, and for parameters such that $\Lambda_{\rm p}$ is much larger than all other scales in the problem, the renormalized effective potential computed with $\Lambda<\Lambda_{\rm p}$ will be essentially the same as the one computed in the limit $\Lambda\rightarrow\infty$. We shall use this simplification when evaluating the effective potential numerically.

\subsection{Geometry of the effective potential}\label{sec:geom}
The geometry of the effective potential is encoded in its field
derivatives. For instance, the curvature for a given value of $\phi$
reads:
\beq\label{eq:g2}
\hat M^2\equiv\frac{\delta^2\gamma}{\delta\phi^2}=m_0^2+\frac{\lambda_0}{2}\,\phi^2
+\frac{\lambda_0}{2}\int_{Q<\Lambda}^T\bar G -\frac{\lambda_0}{2}\,\phi\int_{Q<\Lambda}^T\bar G^2\,\frac{\partial\bar M^2}{\partial\phi}\,,
\eeq
where we made use of the stationnarity condition (\ref{eq:stat}) and $\delta \bar G=-\bar G^2\delta \bar G^{-1}$. From
the flow equation (\ref{eq:flowMphi}), one obtains $\partial\bar M^2/\partial\phi=\bar V\phi$. Plugging this into Eq.~(\ref{eq:g2}) and using Eq.~(\ref{eq:BS}), we arrive at:
\beq\label{eq:g2_bis}
\hat M^2=\bar M^2+\bar V\,\phi^2-\lambda_0\,\phi^2\,.
\eeq
Similarly, the fourth derivative of the effective potential at $\phi=0$ is given by:
\beq\label{eq:g4}
\hat V_{\phi=0}\equiv\left.\frac{\delta^4\gamma}{\delta\phi^4}\right|_{\phi=0}
=\lambda_0-\frac{3}{2}\,\lambda_0\int_{Q<\Lambda}^T\bar G^2_{\phi=0}\left.\frac{\partial^2\bar M^2}{\partial\phi^2}\right|_{\phi=0}\,.
\eeq
The same argument as for $\hat M^2$ leads to:
\beq\label{eq:g4_bis}
\hat V_{\phi=0}=3\bar V_{\phi=0}-2\lambda_0\,.
\eeq
These results illustrate the general discussion given in the Introduction concerning the existence of different definitions of $n$-point functions in a given approximation of the 2PI effective action. For instance, the two-point function (here simply a mass) could be defined either from the solution $\bar M^2$ of the gap equation or from the second derivative $\hat M^2$ of the effective potential. Although these two definitions should coincide for any value of $\phi$ and $T$ in the absence of approximations, Eq.~(\ref{eq:g2_bis}) clearly shows that, in the Hartree approximation, they coincide for $\phi=0$ only. For other approximations, the discrepancy between the different definitions of the two-point function can also be observed for $\phi=0$. Similar remarks apply to the four-point function. Notice however that these discrepancies always appear as higher order effects. In the present case for instance, if we consider a formal expansion in power of the bare coupling $\lambda_0$, we obtain $\hat M^2-\bar M^2=\mathcal{O}(\lambda_0^2)$ as well as $\hat{V}-\bar V={\cal O}(\lambda_0^2)$ which show that the discrepancies are beyond the accuracy of the Hartree approximation.\\

More generally the discrepancies between the different definitions of a given $n$-point function are always of higher order as compared to the accuracy of the 2PI truncation that one considers and are thus of little relevance a priori. However, this is only true for $n\geq 6$. For $n=2$ and $n=4$, the discrepancies between the different definitions of $n$-point functions can be divergent, and they are thus negligeable only after divergences have been properly absorbed. For instance, the two different definitions of the two-point function can bring independent quadratic divergences. In order to remove them, one needs to allow for two independent bare masses following the general prescription of \cite{Berges:2005hc}. It turns out that in the case of the Hartree approximation, the quadratic divergences are not independent but equal\footnote{This has to do with the fact that the two definitions of the mass coincide for $\phi=0$.} which allows one to work with only one bare mass. Nevertheless the sensitivities $\bar M^2$ and $\hat M^2$ with respect to $\Lambda$ are radically different for, as we have seen above, $\bar M^2$ is defined for any value of $\Lambda$ with $\Lambda$-dependent terms of the form $1/\Lambda^2$, whereas, due to the presence of $\lambda_0$ in Eq.~(\ref{eq:g2_bis}), $\hat M^2$ diverges as $\Lambda\rightarrow\Lambda_{\rm p}^-$. For the same reason, the sensitivities of $\bar V$ and $\hat V$ with respect to $\Lambda$ are different. One possible way out is to argue, in a way similar to \cite{AmelinoCamelia:1992nc}, that if one considers $\Lambda\ll\Lambda_{\rm p}$, but much larger than all other scales in the problem, the additional sensitivity of $\hat M^2$ and $\hat V$ as compared to $\bar M^2$ and $\bar V$ is a very mild one. Here we follow a different approach,\footnote{Despite this different point of view, our conclusions regarding the order of the phase transition will be the same as those in \cite{AmelinoCamelia:1992nc}.} in line with the general presentation in \cite{Berges:2005hc}, which is generalizable to higher order approximations. Since the difference of sensitivity with respect to $\Lambda$ comes from the fact that the two definitions $\bar V$ and $\hat V$ of the four-point function differ within a given approximation, it is natural to introduce a bare coupling $\lambda_4$ different from $\lambda_0$ in such a way that the original 2PI functional reads
\beq
\gamma[\phi,G] & = & \frac{m_0^2}{2}\,\phi^2+\frac{\lambda_4}{4!}\,\phi^4+\frac{1}{2}\int_{Q<\Lambda}^T\ln G^{-1}+\frac{1}{2}\int_{Q<\Lambda}^T(Q^2+m_0^2)\,G\nonumber\\
& + & \frac{\lambda_0}{4}\,\phi^2\int^T_{Q<\Lambda} G+\frac{\lambda_0}{8}\left(\int_{Q<\Lambda}^T G\right)^2.
\label{eq:gen_new}
\eeq
Clearly $\lambda_0$ enters the definition of $\bar V$, as before,
whereas $\lambda_4$ appears as a tree level contribution to $\hat
V$. With this modification, using the same steps as those leading to
Eqs.~(\ref{eq:g2_bis}) and (\ref{eq:g4_bis}) we obtain
\beq\label{eq:Mhat_w_l4}
\hat M^2 & = & \bar M^2+\bar V\,\phi^2+\frac{1}{2}(\lambda_4-3\lambda_0)\phi^2\,,\\
\hat V_{\phi=0} & = & 3\bar V_{\phi=0}+(\lambda_4-3\lambda_0)\,.
\eeq
We can now adjust $\lambda_4$ such that $\hat V$ is as insensitive to $\Lambda$ as $\bar V$, that is up to terms of order $1/\Lambda^2$. This however does not fix the finite part of $\lambda_4$. In order to fix the later without introducing additional parameters, and following \cite{Berges:2005hc,Reinosa:2009tc}, we impose the consistency condition:
\beq\label{eq:conscond}
\hat V_{\phi=0,\,T_\star}=\bar V_{\phi=0,\,T_\star}\,,
\eeq
at the renormalization point $\phi=0$ and $T=T_\star$, condition which should hold automatically if no approximation was considered. This condition translates into a particular choice of the bare parameter $\lambda_4$:
\beq\label{eq:l4}
\lambda_4-3\lambda_0=-2\lambda_\star\,.
\eeq
Using this particular choice, the derivatives of the effective potential become then
\beq
\hat M^2 & = & \bar M^2+(\bar V-\lambda_\star)\phi^2\,,\label{eq:curvren}\\
\hat V_{\phi=0} & = & 3\bar V_{\phi=0}-2\lambda_\star\,.
\eeq
Notice that $\hat M^2$ and $\hat V$ are now both cut-off insensitive: the renormalization of $\hat V$ automatically took into account the renormalization of $\hat M^2$. We stress that this is only specific to certain approximations. In general, one would need to introduce a second bare mass which one would fix by imposing another consistency condition. Notice also that, if we expand the bare couplings in powers of the renormalized coupling $\lambda_\star$ using Eqs.~(\ref{eq:l0}) and (\ref{eq:l4}), we find that $\lambda_4-\lambda_0=\mathcal{O}(\lambda^2_\star)$, in line with the fact that the artifacts that the introduction of a second bare coupling $\lambda_4$ is meant to cure lie beyond the accuracy of the Hartree approximation. As a final remark, recall that, as we have shown in the previous section, $\bar M^2\geq m^2_\star>0$ and $\bar V\geq \lambda_\star>0$ at $T=T_\star$ and for any $\phi^2$. It follows that $\hat M^2\geq m_\star^2>0$ at $T=T_\star$ and for any $\phi^2$. This means that the effective potential at the initial temperature $T=T_\star$ is strictly convex.

\subsection{Renormalized effective potential}\label{sec:ren_eff_pot}
We are now in a position to derive a finite expression for the effective potential. Using
Eq.~(\ref{eq:gap}) in Eq.~(\ref{eq:gen_new}) the effective potential
$\gamma(\phi)\equiv\gamma[\phi,\bar G]$ reads
\beq
\gamma(\phi)=\frac{1}{4!}(\lambda_4-3\lambda_0)\phi^4+\frac{\bar M^4-m_0^4}{2\lambda_0}+\frac{1}{2}\int_{Q<\Lambda}^T\ln \bar G^{-1}+\frac{1}{2}\int_{Q<\Lambda}^TQ^2\,\bar G\,.
\eeq
Using Eqs.~(\ref{eq:BS}) and (\ref{eq:l4}), we arrive at
\beq
\gamma(\phi)=-\frac{m_0^4}{2\lambda_0}-\frac{\lambda_\star}{12}\,\phi^4+\frac{\bar M^4}{2\bar V}+\frac{1}{2}\int_{Q<\Lambda}^T\left[\ln \bar G^{-1}+Q^2\,\bar G-\frac{1}{2}\,\bar M^4\,\bar G^2\right]\,.
\eeq
Expanding the integrand for large $Q$ we observe that it has no
quadratic or logarithmic divergence, only a quartic one of the form
$\int_{Q<\Lambda}^T(\ln Q^2+1)$. We can thus write
\beq
\gamma(\phi)=-\frac{m_0^4}{2\lambda_0}-\frac{\lambda_\star}{12}\,\phi^4+\frac{\bar M^4}{2\bar V}+\frac{1}{2}\int_{Q<\Lambda}^T\left[\ln (Q^2+\bar M^2)-\ln Q^2-\bar M^2\,\bar G-\frac{1}{2}\,\bar M^4\,\bar G^2\right]+\frac{1}{2}\int_{Q<\Lambda}^T\Big[\ln Q^2+1\Big]\,.
\eeq
The first integral is convergent. The second is divergent, but one can convince oneself that its
divergence is $T$- and $\phi$-independent. Then, if we subtract the
effective potential evaluated at $T_\star$ and $\phi=0$,
we obtain an explicitly convergent expression
\beq
\label{eq:gamma_final}
\gamma(\phi)-\gamma_{\star}(0)=-\frac{\lambda_\star}{12}\,\phi^4+\frac{\bar M^4}{2\bar V}-\frac{m_\star^4}{2\lambda_\star} & + & \frac{1}{2}\int_{q<\Lambda}\left[\varepsilon_q-q-\frac{\bar M^2}{2\varepsilon_q}-\frac{\bar M^4}{8\varepsilon_q^3}\right]- \frac{1}{2}\int_{q<\Lambda}\left[\varepsilon^\star_q-q-\frac{m_\star^2}{2\varepsilon^\star_q}-\frac{m_\star^4}{8{\varepsilon^\star_q}^3}\right]\nonumber\\
& - & \frac{1}{2}\int_{q<\Lambda} \left[-2T\ln(1-e^{-\varepsilon_q/T})+\bar M^2\,\frac{n_{\varepsilon_q}}{\varepsilon_q}+\bar M^4\,\frac{n_{\varepsilon_q}-\varepsilon_q\,n'_{\varepsilon_q}}{4\varepsilon_q^3}\right]\nonumber\\
& + & \frac{1}{2}\int_{q<\Lambda} \left[-2T_\star\ln(1-e^{-\varepsilon^\star_q/T_\star})+m_\star^2\,\frac{n^\star_{\varepsilon^\star_q}}{\varepsilon^\star_q}+m_\star^4\,\frac{n^\star_{\varepsilon^\star_q}-\varepsilon^\star_q\,{n^\star}'_{\varepsilon^\star_q}}{4{\varepsilon^\star_q}^3}\right]\,,
\eeq
which is our final form for the renormalized effective potential.\footnote{The first two integrals can be done analytically with the
result $(\bar M^4-m_\star^4)/(128\pi^2)$ in the limit $\Lambda\to\infty$.} This formula for the effective potential differs from that in \cite{AmelinoCamelia:1992nc} in that we have used the additional bare parameter $\lambda_4$. The approach we have followed here to obtain a finite effective potential can be generalized to higher orders of approximation, see for instance \cite{Arrizabalaga:2006hj}.

\subsection{Extrema of the effective potential}\label{sec:extrema}
At any temperature, the extrema of the effective potential (\ref{eq:gen_new}) are given by the field equation:
\beq
0=\left.\frac{\delta\gamma}{\delta\phi}\right|_{\bar\phi}\,.
\eeq
In making this equation explicit one can take advantage of the stationarity condition (\ref{eq:stat}). In the Hartree approximation, and taking into account the newly introduced bare coupling $\lambda_4$, we arrive at:
\beq\label{eq:field}
0=\bar\phi\left(m_0^2+\frac{\lambda_4}{6}\,\bar\phi^2+\frac{\lambda_0}{2}\int_{Q<\Lambda}^T\bar G_{\bar\phi}\right)\,.
\eeq
Using the relation (\ref{eq:l4}) as well as the gap equation (\ref{eq:gap}), we
finally arrive at the system of equations:
\beq\label{eq:renfield}
0=\bar\phi\left(\frac{\lambda_\star}{3}\bar\phi^2-\bar M^2_{\bar\phi}\right)  \quad {\rm and} \quad 0=g_\Lambda(\bar M^2_{\bar\phi},\bar\phi^2)\,,
\eeq
where we have made explicit the dependence of $g_\Lambda(M^2)$ with respect to $\phi$. Notice that this system of equations is explicitly finite. This is not surprising for, once the second and fourth derivatives of the effective potential have been renormalized, the effective potential is renormalized (up to an overall constant), and so is any information that can be extracted from it such as the location of its extrema.\footnote{Notice that the way we have obtained an explicitly renormalized field equation differs in spirit from that followed in \cite{Lenaghan:1999si} for the case of the $O(N)$ model. There, the field equation was similar to the one in Eq.~(\ref{eq:renfield}) with $\lambda_\star$ replaced by the bare coupling (denoted by $\lambda$ in Ref.~\cite{Lenaghan:1999si}). Since the latter had no divergence to absorb in the field equation, it was then identified to the renormalized coupling (although this identification was not possible at the level of the gap equation). Even though this result is correct and agrees with Eq.~(\ref{eq:renfield}), its justification in \cite{Lenaghan:1999si} hides the origin of the difficulty, namely the existence of multiple definitions for $n$-point functions and does not allow a generalization to higher orders of approximation. For a discussion of similar issues, see also \cite{Destri:2005qm}.}\\

According to Eq.~(\ref{eq:renfield}), the trivial extremum $\bar\phi=0$ only makes sense as long as the gap equation at $\phi=0$ has a solution. This ceases to be true if $\phi^2_{\rm c}(T)$ is strictly positive, that is when $C_\star<0$ and $T<T_{\rm c}$, see our earlier discussion. On the other hand, the non trivial extrema obey the system:
\beq
0=\frac{\lambda_\star}{3}\bar\phi^2-\bar M^2_{\bar\phi}  \quad {\rm and} \quad 0=g_\Lambda(\bar M^2_{\bar\phi},\bar\phi^2)\,,
\eeq
which we rewrite conveniently as
\beq
\bar\phi^2=\frac{3}{\lambda_\star}\bar M^2_{\bar\phi}  \quad {\rm and} \quad 0=h_\Lambda(\bar M^2_{\bar\phi})\,,
\eeq
where the function $h_\Lambda(M^2)$ does not depend on the field and is given by
\beq\label{eq:h}
h_\Lambda(M^2)\equiv g_\Lambda\left(M^2,\frac{3}{\lambda_\star}M^2\right)=\frac{3}{2}M^2+g_\Lambda(M^2,\phi^2=0)\,,
\eeq
with first and second derivatives equal to
\beq
h'_\Lambda(M^2) & = & \frac{3}{2}+g'_\Lambda(M^2)\,,\\
h''_\Lambda(M^2) & = & g''_\Lambda(M^2)\,.
\eeq
The non trivial extrema of the effective potential can thus be obtained from the zeros of $h_\Lambda(M^2)$, which we discuss below. Notice that, from Eq.~(\ref{eq:curvren}), the curvature of the effective potential at the non-trivial extrema is given by
\beq\label{eq:curvextrema}
\hat M^2=\left(\bar V_{\bar\phi}-\frac{2\lambda_\star}{3}\right)\bar\phi^2=-\lambda_\star\left(\frac{1}{g'_\Lambda(\bar M^2_{\bar\phi})}+\frac{2}{3}\right)\bar\phi^2=-\frac{2\lambda_\star}{3}\frac{h'_\Lambda(\bar M^2_{\bar\phi})}{g'_\Lambda(\bar M^2_{\bar\phi})}\,\bar\phi^2\,.
\eeq
Because $g'_\Lambda(\bar M^2)<0$ when $\Lambda<\Lambda_{\rm p}$, see Appendix~\ref{app:sol}, the sign of the curvature is the sign of $h'_\Lambda(\bar M^2)$.

\subsection{Temperature dependence of the effective potential}\label{sec:temp}
Let us now discuss the zeros of $h_\Lambda(M^2)$ and the corresponding curvature encoded in Eq.~(\ref{eq:curvextrema}) and deduce the shape of the effective potential as we decrease the temperature from $T_\star$ down to $T=0$. Because $h''_\Lambda(M^2)>0$, the function $h'_\Lambda(M^2)$ increases strictly from $h'_\Lambda(0)=-\infty$ to 
\beq
h'_\Lambda(\infty)=\frac{3}{2}+g'_\Lambda(\infty)=\frac{3}{2}-\frac{\lambda_\star}{\lambda_0}=\frac{1}{2}+\frac{\lambda_\star}{2}\int_{Q<\Lambda}^{T_\star}G_\star^2>0\,,
\eeq
where we have used $g'_\Lambda(\infty)=-\lambda_\star/\lambda_0$, see Appendix~\ref{app:sol}, and Eq.~(\ref{eq:l0}). The function $h'_\Lambda(M^2)$ has to vanish for some $\bar M^2_{\rm e}$. It follows that the function $h_\Lambda(M^2)$ has a minimum at $M^2=\bar M^2_{\rm e}$: $h_\Lambda(M^2)$ decreases strictly from $h_\Lambda(0)$ to $h_\Lambda(\bar M^2_{\rm e})$ and then increases towards $h_\Lambda(\infty)=\infty$. The number of non-trivial extrema of the effective potential depends thus on the signs of $h_\Lambda(0)$ and $h_\Lambda(\bar M^2_{\rm e})$ which we now discuss.\\

Let us consider $h_\Lambda(0)$ first. It is nothing but $g_\Lambda(0)$ at $\phi=0$, see Eq.~(\ref{eq:h}), and thus from the discussion above Eq.~(\ref{eq:appp}):
\beq
h_\Lambda(0)=-\frac{\lambda_\star}{2}\,\phi^2_{\rm c}(T,\Lambda)=C_\star+\frac{\lambda_\star}{2}\int_{q<\Lambda}\frac{n_q}{q}\,,
\eeq
where $C_\star$ was defined in Eq.~(\ref{eq:Cstar}) and depends on the
parameters. As we have seen below Eq.~(\ref{eq:Tc}), $\phi^2_{\rm c}(T,\Lambda)$ is strictly negative for $T=T_\star$. It follows that $h_\Lambda(0)$ is strictly positive for $T=T_\star$. Moreover, the previous formula shows that
$h_\Lambda(0)$ decreases strictly as we lower $T$ and reaches
$C_\star$ at $T=0$. If the parameters are such that $C_\star>0$, then
$h_\Lambda(0)$ remains strictly positive down to $T=0$. If
$C_\star=0$, $h_\Lambda(0)$ remains strictly positive down to $T>0$
and vanishes at $T=0$. If $C_\star<0$, $h_\Lambda(0)$ remains strictly
positive while $T>T_{\rm c}$ ($0<T_{\rm c}<T_\star$), vanishes at
$T=T_{\rm c},$ changes sign and remains strictly negative down to $T=0$. The same type of analysis needs to be done for $h_\Lambda(\bar M^2_{\rm e})$. Using Eq.~(\ref{eq:gapr}), we obtain
\beq
h_\Lambda(\bar M^2_{\rm e})=\frac{1}{2}\,\bar M^2_{\rm e}+m^2_\star+\frac{\lambda_\star}{2}\left[\int_{q<\Lambda}\frac{\delta_\star n_{\varepsilon^{\rm e}_q}}{\varepsilon^{\rm e}_q}+(\bar M^2_{\rm e}-m^2_\star)^2\int^{T_\star}_{Q<\Lambda} G_\star^2\,\bar G_{\rm e}\right].
\eeq
Clearly, $h_\Lambda(\bar M_{\rm e})>0$ at $T=T_\star$. Moreover,
because $h'_\Lambda(\bar M_{\rm e})=0$, the thermal dependence of
$h(\bar M_{\rm e})$ is the explicit one, encoded in the tadpole
integral. It follows that $h_\Lambda(\bar M^2_{\rm e})$ decreases as
one decreases $T$. We shall denote by $D_\star$ its value at $T=0$. If
the parameters are such that $D_\star>0$, $h_\Lambda(\bar M^2_{\rm
  e})$ remains strictly positive all the way down to $T=0$. If
$D_\star=0$, it remains strictly positive down to $T>0$ and vanishes
at $T=0$. If $D_\star<0$, $h_\Lambda(\bar M^2_{\rm e})$ remains
strictly positive while $T>T_{\rm s}$ ($0<T_{\rm s}<T_\star$),
vanishes at $T=T_{\rm s},$ changes sign and remains strictly negative down to $T=0.$\\

A priori, discussing the signs of $h_\Lambda(0)$ and $h_\Lambda(\bar M^2_{\rm e})$ involves nine cases, depending on the values of $C_\star$ and $D_\star$, and thus nine regions in parameter space. However, because $C_\star$ and $D_\star$ are the values of $h_\Lambda(0)$ and $h_\Lambda(\bar M^2_{\rm e})$ at $T=0$ and $h_\Lambda(0)>h_\Lambda(\bar M^2_{\rm e})$, we have necessarily $C_\star>D_\star$, which reduces the number of regions to five:\\

\begin{figure}[!tbp]
\begin{center}
\includegraphics[keepaspectratio,width=0.9\textwidth,angle=0]{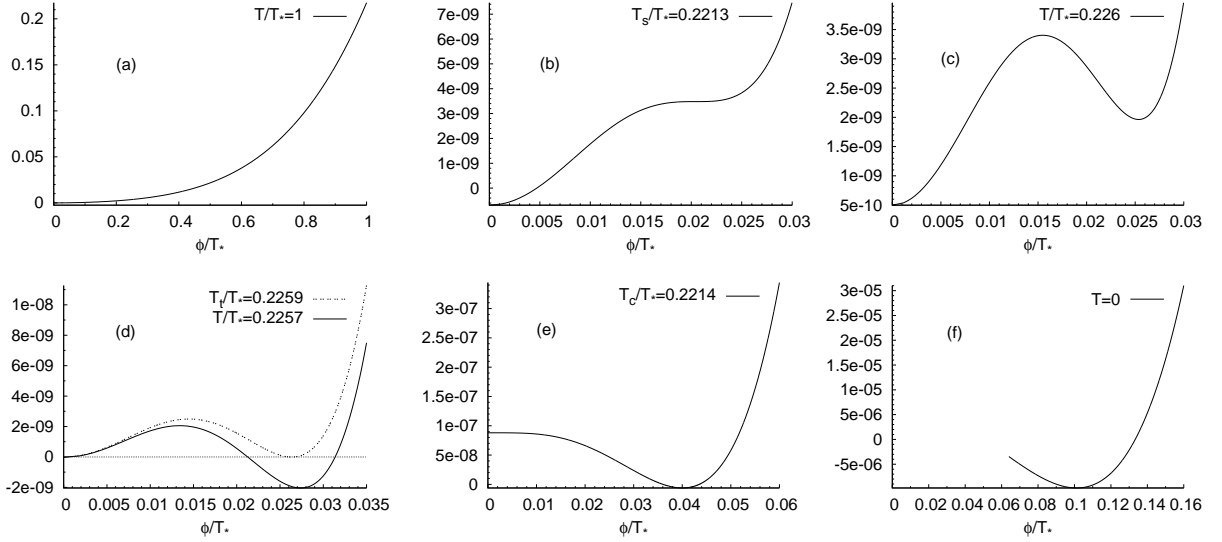}
\caption{The temperature evolution of the effective potential obtained for $m^2_\star/T_\star^2=0.1$, $\lambda_\star=3$ in the limit $\Lambda\rightarrow\infty$ by evaluating Eq.~(\ref{eq:gamma_final}) with $\bar M^2$ calculated
  at $T_\star$ using the $\phi^2$-flow Eq.~(\ref{eq:flowMphi}) and at $T\ne0$
  using the $T$-flow Eq.~(\ref{eq:TflowM}). On each plot, a constant has been subtracted for convenience, without affecting the physical interpretation.
\label{fig:eff_pot}}
\end{center}
\end{figure}

1. Region corresponding to $C_\star<0$: For
such parameters,  $h_\Lambda(\bar M^2_{\rm e})$ changes sign at
$T_{\rm s}$ and $h_\Lambda(0)$ changes sign at $T_{\rm c}<T_{\rm
  s}$. The evolution of the effective potential as we decrease $T$
from $T_\star$ down to $T=0$ goes as follows: 

\begin{itemize}

\item[$-$] For $T_{\rm s}<T\leq T_\star$, both $h_\Lambda(0)$ and $h_\Lambda(\bar M^2_{\rm e})$ are strictly positive and the effective potential has only a trivial extremum at $\phi=0$. It is a minimum because $\hat M^2=\bar M^2>0$ at $\phi=0$. The effective potential is convex, see Fig.~\ref{fig:eff_pot}(a).

\item[$-$] At $T=T_{\rm s}$, $h_\Lambda(0)$ is still strictly positive but $h(\bar M^2_{\rm e})$ vanishes: in addition to the trivial minimum at $\phi=0$, a non trivial extremum appears. According to Eq.~(\ref{eq:curvextrema}), its curvature is zero. The only possibility is that it is an inflection point of the effective potential, see Fig.~\ref{fig:eff_pot}(b).

\item[$-$] For $T_{\rm c}<T<T_{\rm s}$, $h_\Lambda(0)$ is still
  strictly positive and $h(\bar M^2_{\rm e})$ is strictly negative: in
  addition to the trivial minimum at $\phi=0$, two non-trivial extrema
  are present. According to Eq.~(\ref{eq:curvextrema}), the
  non-trivial extremum closest to $\phi=0$ is a maximum and the other
  one is a minimum, see Fig.~\ref{fig:eff_pot}(c). As it will become
  clear in the next point, it turns out that in this temperature interval
  there is a temperature $T_{\rm t}$ ($T_{\rm c}<T_{\rm t}<T_{\rm s}$) 
  at which the minima are degenerate and below which the non trivial minimum
  becomes the absolute minimum, see Fig.~\ref{fig:eff_pot}(d).

\item[$-$] At  $T=T_{\rm c}$, $h_\Lambda(0)$ vanishes and
  $h_\Lambda(\bar M^2_{\rm e})$ is strictly negative: the non-trivial
  extremum which corresponded to a maximum is at $\phi=0$ and merges
  with the trivial minimum. The curvature at $\phi=0$ is zero. Because
  there is still a non-trivial minimum at $\phi\neq 0$, the only
  possibility is that at $T=T_{\rm c}$ the effective potential
  presents a maximum at  $\phi=0$, see Fig.~\ref{fig:eff_pot}(e), with
  the consequence that there was a temperature $T_{\rm t}$ ($T_{\rm c}<T_{\rm t}<T_{\rm s}$) below which the non-trivial minimum became the absolute minimum.

\item[$-$] For $T<T_{\rm c}$, the gap equation has no solution at $\phi=0$: the trivial extremum disappears. The effective potential is defined only for $\phi^2\geq\phi^2_{\rm c}(T)$ and has a single non-trivial minimum, see Fig.~\ref{fig:eff_pot}(f).

\end{itemize}

2. Region corresponding to $C_\star=0$: It is just a line in parameter space. The same analysis as above applies except from the fact that $T_{\rm c}=0$. The final shape of the potential is that of Fig.~\ref{fig:eff_pot}(e).\\

3. Region corresponding to $C_\star>0$ and $D_\star<0$: The same
analysis applies except from the fact that there is no $T_{\rm
  c}$. The final shape of the potential at $T=0$ is either that of Fig.~\ref{fig:eff_pot}(d) or that of Fig.~\ref{fig:eff_pot}(c), but without evaluating the potential we cannot decide whether or not the non-trivial minimum became the absolute one at some temperature.\\
  
4. Region corresponding to $C_\star>0$ and $D_\star=0$: It is just a line in parameter space. The same analysis applies except from the fact that there is no $T_{\rm c}$ and $T_{\rm s}=0$. The final shape of the potential is that of Fig.~\ref{fig:eff_pot}(b).\\

5. Region corresponding to $C_\star>0$ and $D_\star>0$: There are no
$T_{\rm c}$ or $T_{\rm s}$. The final shape of the potential is
convex, see Fig.~\ref{fig:eff_pot}(a).\\

The previous discussion shows that, if there is a phase transition, it is first order due to the finite jump in the location of the absolute minimum of the effective potential (see Fig.~\ref{fig:eff_pot}).  The transition occurs at $T=T_{\rm t}$ while $T_{\rm c}$ and $T_{\rm s}$ correspond to the upper and lower spinodal temperatures at which a maximum and a minimum of the potential merge.\\

\begin{figure}[htbp]
\begin{center}
\includegraphics[keepaspectratio,width=0.6\textwidth,angle=0]{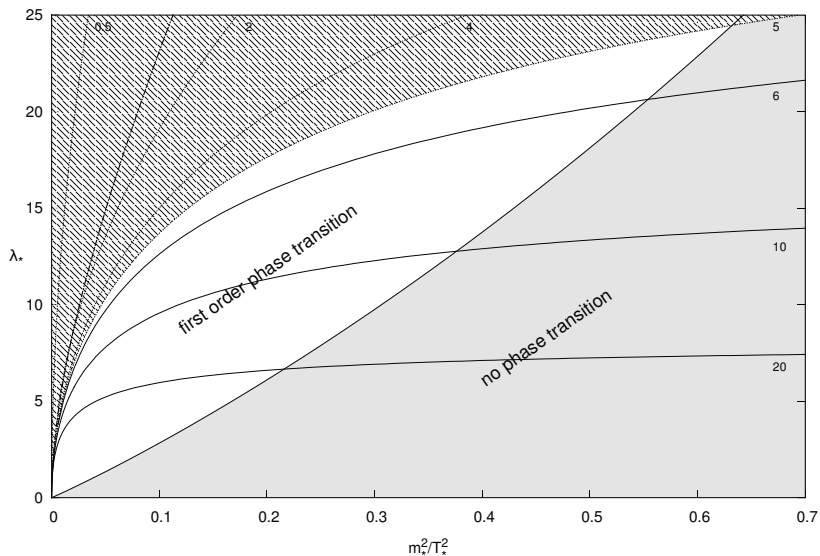}
\caption{Parameter space and order of the phase transition. The labels
  on the curves indicate the value of $\ln(\Lambda_{\rm p}/m_\star)$. We focus on the region $\ln(\Lambda_{\rm p}/m_\star)>5$ which is such our results can be considered cut-off insensitive for a cut-off $\Lambda$ below the Landau scale. Note that this region lies below the dot-dashed curve that is in the region of validity of the condition (\ref{eq:cond_on_RoI}) which played a role in part of our analysis of the gap equation.
\label{fig:params}}
\end{center}
\end{figure}

These conclusions are valid for a given value of the cut-off $\Lambda<\Lambda_{\rm p}$. As we have explained in the previous section, if we want the results to be independent of the cut-off $\Lambda$ we need to choose the parameters such that $\Lambda_{\rm p}\gg m_\star,\,T_\star,\,T$ so that it is possible to choose $\Lambda$ below the Landau scale but larger with respect to all other scales.\footnote{For practical purposes, it is then enough to compute the effective potential for $\Lambda=\infty$ without introducing any substantial difference as compared to the effective potential at $\Lambda<\Lambda_{\rm p}$. One can also perform an analytical study of the effective potential for $\Lambda>\Lambda_{\rm p}$. The only subtlety is that $g'(M^2)$ can be positive for $M^2>\bar M^2_{\rm p}$ and the discussion around Eq.~(\ref{eq:curvextrema}) needs to be revisited. One can convince oneself that the zeros of $h(M^2)$ are all low scales, lying below the high scale $\bar M^2_{\rm p}$, and correspond thus to negative values of $g'(M^2)$.} In what follows we choose $T_\star=1$ which sets the mass unit. Since we are only interested in values of $T$ below $T_\star$, it is then enough to choose $m_\star$ and $\lambda_\star$ such that $\Lambda_{\rm p}/m_\star\gg 1$. For definiteness, we choose to work in the region of parameter space such that $\ln(\Lambda_{\rm p}/m_\star)>5$. Note that, in this regime
\beq
\frac{\bar M^2_{\rm e}(T=0)}{m^2_\star}=\frac{m_\star e}{2\Lambda_{\rm p}}\exp\left(-24\pi^2\int_q\frac{n^\star_{\varepsilon^\star_q}-\varepsilon^\star_q {n^\star}'_{\varepsilon^\star_q}}{2{\varepsilon^\star_q}^3}\right)\ll 1\,,
\eeq
which we obtained from the equation $h'_\Lambda(\bar M^2_{\rm e})=0$ at $T=0$ and for $\Lambda\gg m_\star,\,T_\star$. One can show in the same way that $D_\star$,  which is defined as $h_\Lambda(\bar M^2_{\rm e})$ at $T=0$, is given by
\beq
D_\star=C_\star-\frac{\lambda_\star}{32\pi^2}\bar M^2_{\rm e}(T=0)\,.
\eeq
Since regions 2., 3. and 4. are such that $C_\star\geq 0$ and $D_\star\leq 0$, it follows that $0\leq C_\star\leq\lambda_\star \bar M^2_{\rm e}(T=0)/(32\pi^2)$ and they form a very narrow band, indistinguishable from the line $C_\star=0$ whose equation is\footnote{For practical purposes this curve can be approximated using the high temperature expansion: $\lambda_\star(m_\star)=2m_\star^2/(T_\star^2/12-m_\star T_\star/(8\pi))$.}
\beq\label{eq:Cstar_HTE}
\lambda_\star(m_\star)=2m^2_\star\left[\int_{q}\frac{n^\star_q}{q}-m^4_\star \int^{T_\star}_{Q} \frac{G_\star^2}{Q^2}\right]^{-1}.
\eeq
The parameter space is then essentially divided in two regions, see Fig.~\ref{fig:params}. The lower (grey) region, corresponding to region 5. of the analysis above, is such that no phase transition occurs between $T=T_\star$ and $T=0$. The upper (white) region, corresponding to region 1. of the analysis above, is such that a first order phase transition occurs at $T=T_{\rm t}$. The temperatures $T_{\rm t}$ and $T_{\rm s}$ can only be accessed numerically. We find that for moderate values of the parameters
$\lambda_\star$ and $m^2_\star$ the first order phase transition is
rather weak, that is for reasonable values of the parameters the
difference between $T_{\rm c}$ and $T_{\rm s}$ is small and also for $T_{\rm c}<T<T_{\rm s}$
the non-trivial minimum is close to the trivial one. 
The first order nature of the phase
transition is an artifact of the Hartree approximation which corresponds to a $\mathcal{O}(\lambda)$ truncation of the 2PI functional at skeleton level. Since this effect is rather weak, an improvement of the truncation by including the order-$\lambda^2$ and field-dependent skeleton contribution to the 2PI functional seems to turn the phase transition into second order \cite{Arrizabalaga:2006hj}, in accordance with universality arguments.\\

\begin{figure}[htbp]
\begin{center}
\includegraphics[keepaspectratio,width=0.55\textwidth,angle=0]{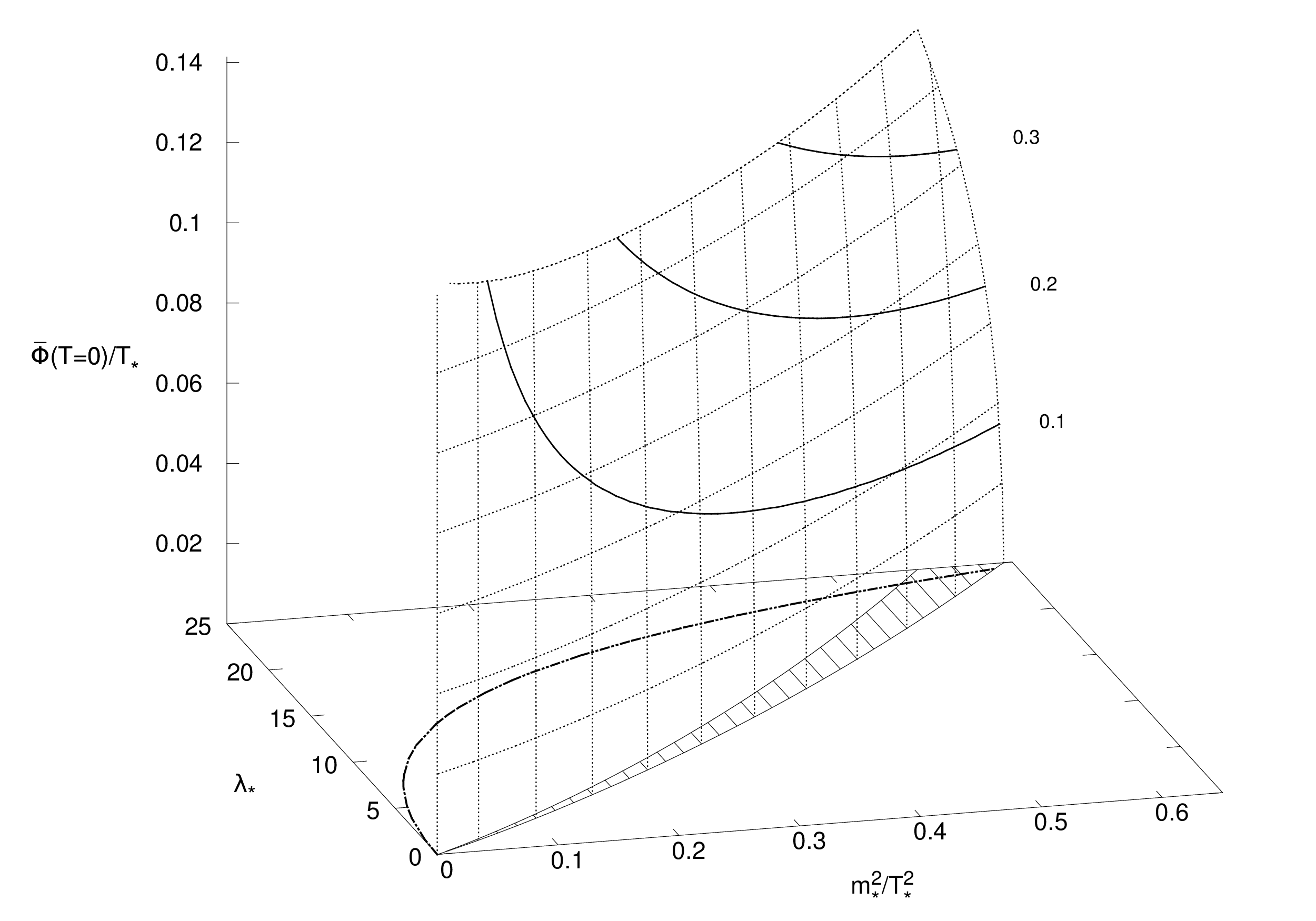}
\caption{ Dependence of the solution of the field equation at $T=0$ on
  the parameters in the region of the parameter space where
  $T_{\rm c}/T_\star \le 0.3$ ($T_c$ increases with $\lambda_\star.$) 
  The labels near the solid curves refers to the solution $\bar M$ of the gap 
  equation at $T=0.$
\label{fig:params_cut}}
\end{center}
\end{figure}

For phenomenological applications a parametrization of the model is
needed, which is usually done at zero temperature. For this reason it
is interesting to know which values of the parameters $m^2_\star$ and $\lambda_\star$
at $T_\star$ are of interest. In the region of the parameter space
where there is a phase transition $T_c$ increases steeply with
$\lambda_\star.$ Therefore, in order for the phase transition to occur at
a reasonable value of temperature compared to $T_\star$ one needs to
restrict the possible values of the parameters in the
$m^2_\star-\lambda_\star$ space. The accessible values of
$\bar\phi(T=0)$ and $\bar M(T=0)$ when the restriction 
$T_c/T_\star \le 0.3$ is imposed can be seen
Fig.~(\ref{fig:params_cut}). This means that the region of the
parameter space which is interesting in practice is a relatively
narrow one, close to the curve given in Eq.~(\ref{eq:Cstar_HTE}).
Setting $T_\star=1$~GeV, these values are in the range of the pion
decay constant, pion masses, and the transition temperature in the
$O(4)$ linear sigma model.

\pagebreak

\section{Conclusions and discussion\label{sec:concl}}

We have studied the phase transition of a real scalar $\varphi^4$ theory in four dimensions at lowest order in the 2PI formalism. Due to the existence of the Landau pole, this model needs to be considered in the presence of a cut-off lying below the scale of the
Landau pole. For values of the parameters such that the Landau pole is much larger than all other scales, renormalization ensure insensitivity of the results with respect to the cut-off scale already below the Landau scale. We have illustrated these questions at the level of the gap equation by investigating under which conditions the cut-off can be continued above the Landau pole to infinity without noticing any sensible change in the solution of the gap equation.  We have
also studied the effective potential and proven analytically that if there is a phase transition, it is 
weakly first order, confirming earlier studies based on a numerical evaluation or a high temperature expansion of the effective potential.\\

Let us end by emphasizing that even though we imposed the renormalization and consistency conditions in the symmetric phase, we could have imposed them in the broken phase as well. Because $\phi=0$ is not always accessible we need to impose these conditions at $\phi=v\neq 0$ (where $v$ is arbitrary for the moment). The bare mass $m_0^2$ and the bare coupling $\lambda_0$ are used to impose the renormalization conditions
\beq
\bar M^2_{\phi=v,\,T=0}=m^2>0 \quad {\rm and} \quad \bar V_{\phi=v,\,T=0}=\lambda>0\,.
\eeq
Unlike what happened at $\phi=0$ in the symmetric phase, the curvature $\hat M^2_{\phi=v,\,T=0}$ is not equal to $\bar M^2_{\phi=v,\,T=0}$. This truncation artifact can be overcome by using the bare coupling $\lambda_4$ to impose the consistency condition:
\beq\label{eq:consist2}
\hat M^2_{\phi=v,\,T=0}=\bar M^2_{\phi=v,\,T=0}\,.
\eeq
Notice then that the consistency condition:
\beq\label{eq:consist4}
\hat V_{\phi=v,\,T=0}=\bar V_{\phi=v,\,T=0}\,,
\eeq
fixes the value of $v$ in terms of $m^2$ and $\lambda$. An explicit calculation leads to:\footnote{This shows in particular that $\lambda<32\pi^2/3\approx 105$.}
\beq
x^2\equiv\frac{v^2}{m^2}=\frac{6}{\lambda-3\lambda^2/(32\pi^2)}\,,
\eeq
One can then check that $v$ cannot be interpreted as the value of $\bar\phi$ (minimum of the potential) at $T=0$ which, given the renormalization and consistency conditions, is such that $\bar\phi^2_{T=0}/m^2=3/\lambda$. Reversely, if one chooses $v$ as the value of $\bar\phi$ at $T=0$, which is the usual choice, then $v^2/m^2=3/\lambda$ and one cannot impose the consistency condition (\ref{eq:consist4}). In this later case, we have checked using similar arguments as above, that the transition cannot be second order.\\
 
The impossibility to impose simultaneously the consistency conditions (\ref{eq:consist2}) and (\ref{eq:consist4}) at the minimum of the potential needs to be regarded as a truncation artifact. It can be cured by exploiting the fact that in higher orders of approximation of the 2PI effective action, two bare masses $m_0^2$ and $m_2^2$ are usually needed to renormalize different quadratic divergences in $\hat M^2$ and $\bar M^2$, see for instance \cite{Arrizabalaga:2006hj}. It is true that in the Hartree approximation these quadratic divergences are equal, but we could still allow for different finite parts to $m^2_0$ and $m_2^2$ and use them to impose both consistency conditions at the minimum of the effective potential. Notice that this approach and the one involving only one bare mass are not completely equivalent. They both agree at leading order in $\lambda$ but differ beyond this order: in the first approach, some of the consistency conditions, which reflect an exact property of the theory are sacrificed, whereas in the second approach, what is sacrificed is the fact that in the exact theory there is only one bare mass. The second approach can be viewed as a renormalization improvement over the first one, for exact properties of the theory are reflected at the level of the renormalized quantities not at the level of the bare ones. It is then interesting to wonder how such an improvement affects the order of the transition. Interestingly, before exploiting the consistency condition (\ref{eq:consist4}) a second order phase transition is possible for parameters such that $v^2/m^2<1/\lambda$. However, the consistency condition (\ref{eq:consist4}) leads once again to a relation between $v^2/m^2$ and $\lambda$:
\beq
0=\frac{\lambda^4}{32\pi^2}\left(\frac{3\lambda}{32\pi^2}-1\right)\frac{v^6}{m^6}+\frac{3\lambda^3}{16\pi^2}\frac{v^4}{m^4}-\lambda \frac{v^2}{m^2}+3\,.
\eeq
For positive $\lambda$, this equation has a positive solution $v^2/m^2$ only for $\lambda<32\pi^2/3$. Moreover, it is possible to show that $v^2/m^2>3/\lambda$ and thus the region $v^2/m^2<1/\lambda$ cannot be accessed. For parameters such that $v^2/m^2>3/\lambda$, it is again possible to convince oneself that the transition cannot be of second order. Then, our conclusions on the nature of the phase transition in the Hartree approximation remain unchanged even within this extended renormalization scheme. The impossibility of a second order phase transition
is an artifact of the Hartree approximation. Therefore, it would be of
interest to see to what extend one could apply the analysis and
methods of the current work to more complicated approximations of the
2PI functional, in which there exist in the literature numerical
evidence for the occurrence of a second order phase transition \cite{Arrizabalaga:2006hj}.
Investigations in this direction are under way.

\acknowledgements{We would like to thank J.-P. Blaizot, G. Fej\H{o}s and J. Serreau for useful discussions on related topics.}

\appendix

\section{Useful identities}\label{app:id}
With a three dimensional and rotation invariant regularization, the tadpole integral can be written as
\beq
\int_{Q<\Lambda}^T \bar G=\int_{q<\Lambda}\frac{1+2\,n_{\varepsilon_q}}{2\,\varepsilon_q}=\int_{Q<\Lambda}^{T_\star}\bar G+\int_{q<\Lambda}\frac{\delta_\star n_{\varepsilon_q}}{\varepsilon_q}\,,\label{eq:id1}
\eeq
with $\varepsilon_q\equiv (q^2+\bar M^2)^{1/2}$, $n_\varepsilon\equiv 1/(e^{\beta\varepsilon}-1)$ and $\delta_\star n_\varepsilon\equiv  n_\varepsilon-n^\star_\varepsilon$, with $n^\star_\varepsilon$ the thermal factor at temperature $T_\star$. The infinitesimal version of Eq.~(\ref{eq:id1}) is also useful and is obtained after taking a derivative with respect to $T$ in Eq.~(\ref{eq:id1}) followed by the limit $T_\star\rightarrow T$. We obtain
\beq
\frac{\partial}{\partial T}\left(\int_{Q<\Lambda}^T\bar G\right)=\int_{Q<\Lambda}^T\frac{\partial\bar G}{\partial T}+\left(\frac{\partial}{\partial T}\int_{Q<\Lambda}^T\right)\bar G\,,
\eeq
with:
\beq
\left(\frac{\partial}{\partial T}\int_{Q<\Lambda}^T\right)\bar G\equiv\int_{q<\Lambda}\frac{\dot{n}_{\varepsilon_q}}{\varepsilon_q}>0\,,
\eeq
and where the dot in $\dot n_\varepsilon$ represents a derivative with respect to the explicit thermal dependence, so that $\dot n_\varepsilon=(\varepsilon/T^2)n_\varepsilon(1+n_\varepsilon)>0$. Similar formulae can be obtained for the bubble integral, for instance:
\beq
\int_{Q<\Lambda}^T\bar G^2=\int_{q<\Lambda}\frac{1+2(n_{\varepsilon_q}-\varepsilon_q n'_{\varepsilon_q})}{4\,\varepsilon_q^3}=\int_{Q<\Lambda}^{T_\star}\bar G^2+\int_{q<\Lambda}\frac{\delta_\star n_{\varepsilon_q}-\varepsilon_q\delta_\star n'_{\varepsilon_q}}{2\,\varepsilon_q^3}\,,\label{eq:id2}
\eeq
where the prime in $n'_\varepsilon$ denotes a derivative with respect to energy, so that $n'_\varepsilon=-(1/T)n_\varepsilon(1+n_\varepsilon)<0$. The infinitesimal form of Eq.~(\ref{eq:id2}) reads
\beq
\frac{\partial}{\partial T}\left(\int_{Q<\Lambda}^T\bar G^2\right)=\int_{Q<\Lambda}^T\frac{\partial\bar G^2}{\partial T}+\left(\frac{\partial}{\partial T}\int_{Q<\Lambda}^T\right)\bar G^2\,,
\eeq
with:
\beq
\left(\frac{\partial}{\partial T}\int_{Q<\Lambda}^T\right)\bar G^2\equiv\int_{q<\Lambda}\frac{\dot{n}_{\varepsilon_q}-\varepsilon_q\dot{n}'_{\varepsilon_q}}{2\,\varepsilon_q^3}>0\,.
\eeq

\section{Solutions of the renormalized gap equation}\label{app:sol}
In order to discuss the solutions of the gap equation $0=g_\Lambda(\bar M^2)$, we study the zeros of
\beq
g_\Lambda(M^2)\equiv -M^2+m^2_\star+\frac{\lambda_\star}{2}\left[\phi^2+\int_{Q<\Lambda}^TG-\int_{Q<\Lambda}^{T_\star}G_\star+(M^2-m^2_\star)\int_{Q<\Lambda}^{T_\star}G_\star^2\right],\label{eq:un}
\eeq
with first and second derivatives given by
\beq
g'_\Lambda(M^2) & = & -1-\frac{\lambda_\star}{2}\left[\int_{Q<\Lambda}^T G^2-\int_{Q<\Lambda}^{T_\star}G_\star^2\right],\label{eq:deux}\\
g''_\Lambda(M^2) & = & \lambda_\star\int_{Q<\Lambda}^T G^3\,.\label{eq:trois}
\eeq
It is useful to notice that $g_\Lambda(M^2)$ and $g'_\Lambda(M^2)$ can be written as
\beq
g_\Lambda(M^2) & = & -\frac{\lambda_\star}{\lambda_0}(M^2-m^2_\star)+\frac{\lambda_\star}{2}\left[\phi^2+\int_{Q<\Lambda}^TG-\int_{Q<\Lambda}^{T_\star}G_\star\right],\label{eq:quatre}\\
g'_\Lambda(M^2) & = & -\lambda_\star\left[\frac{1}{\lambda_0}+\frac{1}{2}\int_{Q<\Lambda}^T G^2\right].\label{eq:cinq}
\eeq
We shall also use the explicitly renormalized form of $g_\Lambda(M^2)$
given in Eq.~(\ref{eq:gapr}). As explained at the beginning of
Section~\ref{ss:UVity}, we perform our analysis in the regime of
interest, that is where $\Lambda$ and $\Lambda_{\rm p}$ are much
larger than all the other scales $T_\star,\,m_\star,\,\phi,$ and $T$. Moreover, we are only interested in positive solutions $\bar M^2$ of the gap equation.

\subsection{Solution for $\Lambda<\Lambda_{\rm p}$}
The function $g''_\Lambda(M^2)$ is strictly positive, see
Eq.~(\ref{eq:trois}). It follows that $g'_\Lambda(M^2)$ increases
strictly from $g'_\Lambda(0)=-\infty$ (the bubble integral in
Eq.~(\ref{eq:cinq}) diverges positively as $M^2\rightarrow 0$) to
$g'_\Lambda(\infty)=-\lambda_\star/\lambda_0$ (the bubble integral in
Eq.~(\ref{eq:cinq}) is suppressed as $M^2\gg\Lambda^2,\,T^2$). Because
$\Lambda<\Lambda_{\rm p}$, it follows from Eq.~(\ref{eq:l0}) and the
  definition of $\Lambda_{\rm p}$ given in Eq.~(\ref{eq:Landau}) that $\lambda_0>0$ and
therefore $g'_\Lambda(\infty)<0$. In consequence $g'_\Lambda(M^2)<0$ and the function $g_\Lambda(M^2)$ decreases strictly from $g_\Lambda(0)\equiv\lambda_\star(\phi^2-\phi^2_{\rm c}(T,\Lambda))/2$ where
\beq\label{eq:appp}
\phi^2_{\rm c}(T,\Lambda)=-\frac{2\,m^2_\star}{\lambda_\star}-\int_{Q<\Lambda}^T\frac{1}{Q^2}+\int_{Q<\Lambda}^{T_\star}G_\star+m^2_\star\int_{Q<\Lambda}^{T_\star}G_\star^2\,,
\eeq
to $g_\Lambda(\infty)=-\infty$ (the tadpole integral in Eq.~(\ref{eq:quatre}) is suppressed as $M^2\gg\Lambda^2,\,T^2$). Then, the existence of a solution to the gap equation depends on the sign of $g_\Lambda(0)$ or, in other words, on the value of $\phi^2$ as compared to $\phi^2_{\rm c}(T,\Lambda)$: if $\phi^2>\phi^2_{\rm c}(T,\Lambda)$, the gap equation admits a unique and strictly positive solution; if $\phi^2=\phi^2_{\rm c}(T,\Lambda)$, the unique solution to the gap equation is equal to zero; if $\phi^2<\phi^2_{\rm c}(T,\Lambda)$, the gap equation has no solution.\\

Notice that, so far, $\phi^2_{\rm c}(T,\Lambda)$ does not need to be
positive. In practice however, we shall consider only positive values
of $\phi^2$ and it is useful to determine when $\phi^2_{\rm
  c}(T,\Lambda)$ is positive as well. Using the explicitly convergent form of $g_\Lambda(M^2)$ given in Eq.~(\ref{eq:gapr}), it is convenient to write
\beq
\phi^2_{\rm c}(T,\Lambda)=-\frac{2\,C_\star}{\lambda_\star}-\int_{q<\Lambda}\frac{n_q}{q}\,,
\eeq
with
\beq
C_\star\equiv m^2_\star+\frac{\lambda_\star}{2}\left[-\int_{q<\Lambda}\frac{n^\star_q}{q}+m^4_\star \int^{T_\star}_{Q<\Lambda} \frac{G_\star^2}{Q^2}\right].
\eeq
If the parameters $\Lambda$, $T_\star$, $m_\star$, and $\lambda_\star$
are such that $C_\star \geq 0$, it is clear that $\phi^2_{\rm
  c}(T,\Lambda)$ is negative for any value of $T$ and therefore, it is of no relevance in practice because the gap equation admits a solution for any value of $\phi^2\geq 0$ and $T$. On the contrary, when $C_\star<0$, then $\phi^2_{\rm c}(T,\Lambda)$ becomes strictly positive for $T<T_{\rm c}$ with the ``critical'' temperature $T_{\rm c}$ defined by
\beq
\int_{q<\Lambda}\frac{n^{\rm c}_q}{q}=-\frac{2C_\star}{\lambda_\star}\,.
\eeq
In this case, $\phi^2_{\rm c}(T,\Lambda)$ becomes relevant because the gap equation admits a (positive) solution only for $\phi^2\geq\phi^2_{\rm c}(T,\Lambda)$. Moreover, in the regime of interest, one can neglect the dependence of $C_\star$ and of the thermal integrals with respect to $\Lambda$. The critical value $\phi^2_{\rm c}(T,\Lambda)$ is then independent of $\Lambda$ and one has $\phi^2_{\rm c}(T)=(T^2_{\rm c}-T^2)/12$ with $T^2_{\rm c}=-24C_\star/\lambda_\star$.

\subsection{Solution for $\Lambda=\Lambda_{\rm p}$}
From the definition of $\Lambda_{\rm p}$ in Eq.~(\ref{eq:Landau}), this case corresponds formally to $1/\lambda_0=0$. The function $g'_\Lambda(M^2)$ increases strictly from $g'_\Lambda(0)=-\infty$ to $g'_\Lambda(\infty)=-\lambda_\star/\lambda_0=0$. It follows that $g'_\Lambda(M^2)<0$ and the function $g_\Lambda(M^2)$ decreases strictly from $g_\Lambda(0)\equiv\lambda_\star(\phi^2-\phi^2_{\rm c}(T))/2$ to $g_\Lambda(\infty)\equiv\lambda_\star(\phi^2-\phi^2_{\rm p})/2$ (the tadpole integral in Eq.~(\ref{eq:quatre}) is suppressed as $M^2\gg\Lambda^2$), with
\beq
\phi^2_{\rm p}=\int_{Q<\Lambda_{\rm p}}^{T_\star} G_\star>0\,.
\eeq
The existence of a solution to the gap equation now depends on the signs of $g_\Lambda(0)$ and $g_\Lambda(\infty)$ or, in other words, on the value of $\phi^2$ as compared to $\phi^2_{\rm c}(T)$ an $\phi^2_{\rm p}$. Notice that, because $g_\Lambda(0)>g_\Lambda(\infty)$, we have $\phi^2_{\rm p}>\phi^2_{\rm c}(T)$. We can thus distinguish three different regimes: if $\phi^2\geq\phi^2_{\rm p}$, the gap equation has no solution; if $\phi^2_{\rm p}>\phi^2\geq\phi^2_{\rm c}(T)$, the gap equation admits a unique solution, which vanishes for $\phi^2=\phi^2_{\rm c}(T)$; if  $\phi^2<\phi^2_{\rm c}(T)$, the gap equation has again no solution. Notice that, in the regime of interest, $\phi^2_{\rm p}\sim\Lambda^2_{\rm p}/(8\pi^2)$ is a large scale.\\

It is important to stress that the behavior of the solution $\bar M^2$ of the gap equation as $\Lambda\rightarrow\Lambda_{\rm p}^-$ is rather different depending on the value of $\phi^2$. To see this, note that for a given $\mu$, the limit of $g_\Lambda(\mu^2)$ as $\Lambda\rightarrow\Lambda_{\rm p}^-$ is $\lambda_\star(\phi^2-\phi^2_{\rm p}+\int_{Q<\Lambda_{\rm p}}^TG_\mu)/2$, see Eq.~(\ref{eq:quatre}). Clearly, if $\phi^2\geq\phi^2_{\rm p}$, this limit is strictly positive for any $\mu$, which means that for any $\mu$, there exists a value of $\Lambda<\Lambda_{\rm p}$ above which $g_\Lambda(\mu^2)\geq 0$ and thus $\bar M^2\geq\mu^2$. This shows that the solution of the gap equation diverges as $\Lambda\rightarrow\Lambda_{\rm p}^-$ when $\phi^2\geq\phi^2_{\rm p}$. If $\phi^2<\phi^2_{\rm p}$, nothing of this kind happens and the solution can be extended beyond the Landau scale. We now study what happens beyond this scale.

\subsection{Solutions for $\Lambda>\Lambda_{\rm p}$}
This case corresponds to $\lambda_0<0$. The function $g'_\Lambda(M^2)$
increases strictly from $g'_\Lambda(0)=-\infty$ to
$g'_\Lambda(\infty)=-\lambda_\star/\lambda_0>0$. It follows that
$g'_\Lambda(M^2)$ vanishes for some $\bar M^2_{\rm p}(T,\Lambda)$
(in view of Eq.~(\ref{eq:cinq}) this mass does not depend on $\phi^2$). Then, the function $g_\Lambda(\bar M^2)$ has a minimum at $M^2=\bar M^2_{\rm p}(T,\Lambda)$: it decreases strictly from $g_\Lambda(0)\equiv\lambda_\star(\phi^2-\phi^2_{\rm c}(T))/2$ to $g_\Lambda(\bar M_{\rm p}^2(T,\Lambda))\equiv\lambda_\star(\phi^2-\phi^2_{\rm p}(T,\Lambda))/2$ and then increases towards $g_\Lambda(\infty)=\infty$ (the tadpole integral in Eq.~(\ref{eq:quatre}) is suppressed as $M^2\gg\Lambda^2$). The existence of solutions to the gap equation depends now on the signs of $g_\Lambda(0)$ and $g_\Lambda(\bar M^2_{\rm p}(T,\Lambda))$ or, in other words, on the value of $\phi^2$ as compared to $\phi^2_{\rm c}(T)$ and $\phi^2_{\rm p}(T,\Lambda)$. Notice that because $g_\Lambda(0)>g_\Lambda(\bar M^2_{\rm p}(T,\Lambda))$, we have $\phi^2_{\rm p}(T,\Lambda)>\phi^2_{\rm c}(T)$. We can thus distinguish four different regimes concerning the number of solutions to the gap equation: if $\phi^2>\phi^2_{\rm p}(T,\Lambda)$, there is no solution; if $\phi^2=\phi^2_{\rm p}(T,\Lambda)$, there is one solution, equal to $\bar M^2_{\rm p}(T,\Lambda)$; if $\phi^2_{\rm p}(T,\Lambda)>\phi^2\geq \phi^2_{\rm c}(T)$, there are two solutions, one smaller than $\bar M^2_{\rm p}(T,\Lambda)$, the other larger than $\bar M^2_{\rm p}(T,\Lambda)$; if $\phi^2<\phi^2_{\rm c}(T)$, there is again one solution, larger than $\bar M^2_{\rm p}(T,\Lambda)$.\\

It is useful to study the dependence of $\bar M^2_{\rm p}(T,\Lambda)$
and $\phi^2_{\rm p}(T,\Lambda)$ on $\Lambda$. Let us consider $\bar M^2_{\rm p}(T,\Lambda)$ first. It is
defined by $g'_\Lambda(\bar M^2_{\rm p}(T,\Lambda))=0$. Using
Eq.~(\ref{eq:cinq}), it follows already that $\bar M^2_{\rm
  p}(T,\Lambda)\rightarrow\infty$ as $\Lambda\rightarrow\Lambda^+_{\rm
  p},$ because $-1/\lambda_0\rightarrow 0^+$. Let us now show that, in the regime of interest, $\bar M^2_{\rm p}(T,\Lambda)>m^2_\star$ or, in other words that $g'_\Lambda(m^2_\star)<0$. From Eq.~(\ref{eq:deux}), we write
\beq
g'_\Lambda(m^2_\star) & = & -1-\frac{\lambda_\star}{2}\left[\int_{Q<\Lambda}^T G^2_\star-\int_{Q<\Lambda}^{T_\star}G_\star^2\right]\nonumber\\
& \leq & -1-\frac{\lambda_\star}{2}\left[\int_{Q<\Lambda}^{T=0} G^2_\star-\int_{Q<\Lambda}^{T_\star}G_\star^2\right]\nonumber\\
& \leq & -1+\frac{\lambda_\star}{2}\int_q\frac{n^\star_{\varepsilon^\star_q}-\varepsilon^\star_q{n^\star}'_{\varepsilon^\star_q}}{2{\varepsilon^\star_q}^3}\,,
\eeq
where we have used Eq.~(\ref{eq:id2}) and the fact that the bubble integral increases with $T$ and $\Lambda$. Then $g'_\Lambda(m^2_\star)<0$ if we require that
\beq\label{eq:cond_on_RoI}
\frac{2}{\lambda_\star}>\int_q\frac{n^\star_{\varepsilon^\star_q}-\varepsilon^\star_q{n^\star}'_{\varepsilon^\star_q}}{2{\varepsilon^\star_q}^3}\,.
\eeq
In terms of the Landau pole this condition reads
\beq
\frac{1}{16\pi^2}\left[{\rm Arcsinh}\left(\frac{\Lambda_{\rm p}}{m_\star}\right)-\frac{\Lambda_{\rm p}}{\sqrt{\Lambda^2_{\rm p}+m^2_\star}}\right]\geq\int_{q>\Lambda_{\rm p}}\frac{n^\star_{\varepsilon^\star_q}-\varepsilon^\star_q{n^\star}'_{\varepsilon^\star_q}}{4{\varepsilon^\star_q}^3}\,,
\eeq
which is clearly fulfilled in the regime of interest. Finally, in order to study the $\Lambda$-dependence of $\bar M^2_{\rm p}(T,\Lambda)$, let us take a derivative with respect to $\Lambda$ in the defining equation $0=g'_\Lambda(\bar M^2_{\rm p}(T,\Lambda))$. We arrive at
\beq
\frac{\partial\bar M^2_{\rm p}}{\partial\Lambda}=-\frac{1}{g''_\Lambda(\bar M^2_{\rm p})}\frac{\partial g'_\Lambda}{\partial\Lambda}\,.
\eeq
Using Eqs. (\ref{eq:deux}) and (\ref{eq:id2}), this reads more explicitly:
\beq
\frac{\partial\bar M^2_{\rm p}}{\partial\Lambda}=\frac{\lambda_\star}{2g''_\Lambda(\bar M^2_{\rm p})}\frac{\Lambda^2}{2\pi^2}\left[\frac{1+2(n_{\varepsilon^{\rm p}_\Lambda}-\varepsilon^{\rm p}_\Lambda n'_{\varepsilon^{\rm p}_\Lambda})}{4{\varepsilon^{\rm p}_\Lambda}^3}-\frac{1+2(n^\star_{\varepsilon^\star_\Lambda}-\varepsilon^\star_\Lambda {n^\star}'_{\varepsilon^\star_\Lambda})}{4{\varepsilon^\star_\Lambda}^3}\right].
\eeq
Now, because $T\leq T_\star$ and $\bar M^2_{\rm p}(T,\Lambda)>m^2_\star$, the right-hand-side of this last equation is strictly negative and $\bar M^2_{\rm p}(T,\Lambda)$ decreases strictly with $\Lambda$. Because it is bounded from below, it has a limiting value $\bar M^2_{\rm p}(T,\infty)$ which one could determine from Eq.~(\ref{eq:deux}). In the regime of interest, one obtains $\bar M^2_{\rm p}(T,\infty)\sim4\Lambda^2_{\rm p}/e^2$.\\

Consider finally $\phi^2_{\rm p}(T,\Lambda)$ which is defined as the particular value of $\phi^2$ such that $g_\Lambda(\bar M^2_{\rm p}(T,\Lambda))$ is equal to zero. Making the field dependence of $g_\Lambda(M^2)$ explicit we can then write
\beq
0=g_\Lambda(\bar M_{\rm p}^2(T,\Lambda),\phi^2_{\rm p}(T,\Lambda))\,.
\eeq
Taking a field derivative with respect to $\Lambda$ and using the fact that $g'_\Lambda(\bar M^2_{\rm p}(T,\Lambda))=0$, we obtain:
\beq
0=\frac{\partial g_\Lambda}{\partial\phi^2}\frac{\partial\phi^2_{\rm p}}{\partial\Lambda}+\frac{\partial g_\Lambda}{\partial \Lambda}\,.
\eeq
A straightforward calculation using the explicitly convergent form of $g_\Lambda(M^2)$ given in Eq.~(\ref{eq:gapr}) leads to:
\beq
\frac{\partial\phi^2_{\rm p}}{\partial\Lambda} & = & -\frac{\Lambda^2}{2\pi^2}\,(\bar M^2_{\rm p}-m^2_\star)^2\int_{q_4} G_\star^2\,\bar G_{\rm p}+{\rm thermal}\,,
\eeq
where $\bar G_{\rm p}\equiv 1/(Q^2+\bar M^2_{\rm p})$, the propagators are evaluated for $q^2=\Lambda^2$ and $q_4$ denotes the continuous Euclidean frequency at $T=0$. In the regime of interest $\bar M^2_{\rm p}(T,\Lambda)>\bar M^2_{\rm p}(T,\infty)\gg m^2_\star$. One can thus neglect the thermal contribution in the previous formula. It follows that $\phi^2_{\rm p}(T,\Lambda)$ decreases strictly with $\Lambda$. For $\Lambda\rightarrow\Lambda^+_{\rm p}$, the minimum $\bar M^2_{\rm p}(T,\Lambda)$ of $g_\Lambda(M^2)$ is sent to $\infty$. Thus $g_\Lambda(\bar M^2_{\rm p})=\lambda_\star(\phi^2-\phi^2_{\rm p}(T,\Lambda))/2\rightarrow g_{\Lambda_{\rm p}}(\infty)=\lambda_\star(\phi^2-\phi^2_{\rm p})/2$ and thus $\phi^2_{\rm p}(\Lambda^+_{\rm p})=\phi^2_{\rm p}$. Moreover, using Eqs.~(\ref{eq:un}) and (\ref{eq:deux}) with $g'_\infty(\bar M^2_{\rm p}(\infty))=0$, we obtain
\beq
\phi^2_{\rm p}(\infty)=-\int_q\frac{\delta_\star n_{\varepsilon^\star_q}}{\varepsilon^\star_q}+(\bar M^2_{\rm p}(\infty)-m^2_\star)^2\int_Q^T\bar G_{\rm p}^2(\infty) G_\star\,,
\eeq
which is strictly positive for $T\leq T_\star$. In the regime of interest, only the zero temperature contribution of the second term dominates and we find
\beq
\phi^2_{\rm p}(\infty)\sim \frac{\bar M^2_{\rm p}(\infty)}{8\pi^2}\int_0^\infty dQ\,\frac{Q^3}{(Q^2+1)^2(Q^2+m^2/\bar M^2_{\rm p}(\infty))}\sim\frac{\Lambda^2_{\rm p}}{4\pi^2e^2}\,.
\eeq
Notice finally that :
\beq
\frac{\partial\bar M^2}{\partial\phi^2}=-\frac{\lambda}{2}\frac{1}{g'_\Lambda(\bar M^2)}\,.
\eeq
This shows that, as one increases $\phi^2$, one of the solutions increases while the other decreases. They both become equal to $\bar M^2_{\rm p}(T,\Lambda)$ when $\phi^2=\phi^2_{\rm p}(T,\Lambda)$. This applies in particular to the case $\Lambda=\infty$.

\end{document}